\documentclass[12pt]{article}
\pdfoutput=1
\usepackage{fullpage,amsmath,amssymb,graphicx,mathrsfs}
\usepackage{xcolor}
\usepackage[sc]{titlesec}

\title{Consistent truncations of supergravity and 1/2-BPS RG flows in $4d$ SCFTs}
\author{}
\date{}
\newcommand{\dd}{\mathrm{d}}
\newcommand{\ee}{\mathrm{e}}

\newcommand{\as}{\epsilon}
\newcommand{\wj}{\wedge}
\newcommand{\hs}{\star\,}
\newcommand{\hf}{\ast}

\DeclareMathOperator{\arcsinh}{arcsinh}

\numberwithin{equation}{section}

\begin{document}

\begin{titlepage}

\vfill

\begin{flushright}
ICCUB-20-006\\
\end{flushright}

\vfill

\begin{center}
   \baselineskip=16pt
   {\LARGE\textsc{Consistent truncations of supergravity and $1/2$-BPS RG flows in $4d$ SCFTs}}
  \vskip 1.5cm
Ant\'on F. Faedo$^1$, Carlos Nunez$^2$ and Christopher Rosen$^1$\\
     \vskip .6cm
             \begin{small}\vskip .6cm
      \textit{$^1$Departament de F\'\i sica Qu\`antica i Astrof\'\i sica and Institut de Ci\`encies del Cosmos (ICC),\\  Universitat de Barcelona, Mart\'\i\  i Franqu\`es 1, ES-08028, Barcelona, Spain}
        \end{small}\\
             \begin{small}\vskip .6cm
      \textit{$^2$Department of Physics, Swansea University, Swansea SA2 8PP, United Kingdom}
        \end{small}\\
                               \end{center}
\vfill

\begin{center}
\textbf{Abstract}
\end{center}
\begin{quote}
With the purpose of holographically describing flows from a large family of  four dimensional ${\cal N}=1$ and ${\cal N}=2$ conformal field theories, we discuss truncations of seven dimensional supergravity to five dimensions. We write explicitly the reduced gauged supergravity and find BPS equations for simple configurations. Lifting these flows to eleven dimensions or Massive IIA supergravity, we present string duals to RG flows from strongly coupled conformal theories when deformed by marginal and/or relevant operators. We further discuss observables common to infinite families of ${\cal N}=1$ and ${\cal N}=2$ QFTs in this context.
\end{quote}

\vfill
\begin{center}
\textit{Dedicated to the memory of our extraordinary friend Steve Gubser, whose enthusiasm both at the blackboard and in the mountains will inspire us always.}
\end{center}

\vfill

\end{titlepage}


\section{Introduction}


An important by-product of the Maldacena conjecture \cite{Maldacena:1997re}, is the study of supersymmetric and conformal field theories (SCFTs) in diverse dimensions\footnote{Here we mostly refer to SCFTs in diverse dimensions. It is important to keep in mind that the studies on SCFTs in four dimensions originated in \cite{Sohnius:1981sn}, where many interesting perturbative and non-perturbative results were derived.}.
The last two decades witnessed  a large effort  in the classification of Type II or M-theory backgrounds with $AdS_d$-factors, see for example \cite{Gauntlett:2004zh},\cite{Gutowski:2014ova}. These solutions are conjectured to be dual to  Conformal Field Theories (CFTs) in dimension $d-1$ and with different amounts of Supersymmetry (SUSY). In the case in which the backgrounds preserve eight Poincar\'e supercharges major progress has been achieved.

In fact, for the case of ${\cal N}=2$ superconformal field theories in four dimensions, the field theories studied in \cite{Gaiotto:2009we} have holographic duals first discussed in \cite{Gaiotto:2009gz}, and further elaborated on in \cite{ReidEdwards:2010qs}-\cite{Bah:2019jts} (among other places). The case of five-dimensional SCFTs was analysed from the field-theoretical and holographic viewpoints in \cite{DHoker:2016ujz}-\cite{Bergman:2018hin},  as well as in other interesting works.   An infinite family of six-dimensional ${\cal N}=(1,0)$ SCFTs was discussed from both the field-theoretical and holographic points of view in \cite{Apruzzi:2015wna}-\cite{Hanany:1997gh}. For three-dimensional  ${\cal N}=4$ SCFTs, the field theories presented in \cite{Gaiotto:2008ak} were discussed holographically early on, in e.g. \cite{DHoker:2007hhe}-\cite{Lozano:2016wrs}. Two-dimensional SCFTs  and their $AdS$ duals have also received much attention. From a holographic perspective, recent work in this direction can be found in \cite{Lozano:2019emq}-\cite{Couzens:2019mkh}.

The generic string or M-theory backgrounds dual to CFTs with four Poincar\'e supercharges are harder to classify. In the case of four-dimensional SCFTs one can find some efforts along this line in \cite{Gauntlett:2004zh}, \cite{Bah:2015fwa}.

In the holographic approach, after identifying these diverse $AdS$ solutions with their dual CFTs, the most natural step is to study the dynamics that ensues by deforming the CFT and flowing away from the fixed-point theory. The Holographic Renormalisation Group (H-RG), a set of ideas developed in \cite{deBoer:1999tgo}-\cite{Papadimitriou:2004ap}, is a well studied technical tool that allows a geometric understanding of many details of these flows. In particular, it can be used to extract from a given gravitational background the field theory sources and vacuum expectation values for operators involved in the deformation of the fixed-point theory.

One of the primary strengths of the Renormalisation Group, holographic or otherwise, is its versatility. It has been successfully employed to uncover the low energy properties of theories with and without Lagrangian descriptions, in theories whose coupling is weak or strong, and for theories which enjoy varying amounts of (super)symmetry. A particularly interesting arena, in which many of these ideas intersect, is in the study of RG ``flows across dimensions". This refers to deformations of a UV theory in dimension $d+n$, which drive the theory to an IR fixed point whose physics is effectively $d$-dimensional.

An important class of such flows arises when the deformation driving the flow corresponds to placing the $(d+n)$-dimensional UV theory on a $n$-dimensional manifold. In its earliest incarnation, this is of course the scenario of Kaluza and Klein. More recently, this idea has been rejuvenated and exploited with great success, notably in the study of the landscape of interacting superconformal field theories in four dimensions \cite{Maldacena:2000mw, Benini:2009mz, Bah:2012dg}. 

Maldacena's conjectured duality offers a powerful perspective on these ideas, realising the RG flow geometrically. This perspective was taken in \cite{Maldacena:2000mw}, in which supergravity solutions corresponding to wrapping a large number of $M5$-branes on a Riemann surface were constructed. These solutions preserve either four or eight Poincar\'e supercharges, and were interpreted holographically as M-theory solutions dual to four-dimensional SCFTs obtained by placing the six-dimensional (2,0) $M5$-brane theory on a Riemann surface with a topological twist. 
In the subsequent years, starting with \cite{Gaiotto:2009we, Benini:2009mz}, substantial work has been devoted to understanding the properties of these four-dimensional field theories.

 A primary motivation of this present work is to continue this exploration, returning to the gravitational picture provided by holographic duality. To make progress, we will appeal to two consistent truncations of seven-dimensional supergravity with known uplifts to ten and/or eleven-dimensional supergravity. The truncations result in five-dimensional gauged supergravity theories with $\mathcal{N}=2$ and $\mathcal{N}=4$ supersymmetry whose solutions include those of \cite{Maldacena:2000mw}.

Importantly, the five-dimensional perspective grants access to sectors of these theories which consistently decouple from the rest of the dynamics. In particular, our approach allows us to identify and study a subset of supersymmetric states and RG flows in these theories without the necessity of confronting the computational difficulties of working with higher-dimensional theories of supergravity. An interesting feature of our approach is that a particular solution to the five-dimensional supergravity equations of motion may have more than one uplift to higher dimensions. In this sense, the flows we find are capturing universal Physics in infinite classes of outwardly distinct four-dimensional SCFTs. This universality in the behaviour of some physical phenomena was also pointed out in \cite{Bobev:2017uzs}.

The outline of our work is as follows: in section \ref{sec:Nis2Theory}, we introduce a consistent truncation of seven-dimensional minimal gauged supergravity. The reduction ansatz incorporates a topological twist, marrying certain gauge fields in the supergravity theory to the spin connection on a Riemann surface. The bosonic sector of the reduced theory is that of a $\mathcal{N}=2$ gauged supergravity in five dimensions, with gauge group $SO(2)\times \mathbb{R}$. 

In section \ref{sec:Nis1flos}, borrowing the results of \cite{Ceresole:2001wi}, we use the data of this supergravity theory to obtain a set of BPS equations governing the supersymmetric flows which preserve four supercharges. Although the equations can be studied for any value of the curvature of the Riemann surfaces, we focus on the particular case in which this surface is (a quotient of) $H^2$. In this case, we can holographically identify supersymmetric solutions corresponding to supersymmetric states in the dual SCFT. 

In section \ref{sec:10n11dUp} we reproduce known uplift formulae which allow us to embed the solutions to our five-dimensional theory in ten and eleven dimensions. The content of this section is not new, but crucial for understanding the higher-dimensional features of our solutions. As an example, we exhibit explicitly the admissibility criterion of \cite{Maldacena:2000mw} for the singularities which arise in the IR of our five-dimensional flows. Some of our solutions end in an ``admissible" singularity, and we prefer to think about these solutions as a first step towards the holographic description of the QFT dynamics. It is plausible that a different (perhaps more elaborate) truncation to gauged supergravity leads to a resolution of such singular behaviours.

Section \ref{sec:Nis4} follows the same lines of inquiry, but this time in a very recently discovered truncation of {\it maximal} gauged supergravity in seven dimensions \cite{Cheung:2019pge}. We study supersymmetric solutions in this theory which preserve at least eight supercharges. Interestingly, when the Riemann surface is taken to be $H^2$ or a quotient thereof, we succeed in finding an analytic solution which generalises a solution presented in \cite{Bobev:2019ylk} in the holographic study of the $\Omega$--background. For certain values of the solution's parameters, which correspond to vacuum expectation values for operators in the dual field theory, the dual state shares some important features with a well known Coulomb branch solution of $\mathcal{N}=4$ Super Yang-Mills \cite{Freedman:1999gk}.

In section \ref{sec:QFT} we highlight some lessons we can extract holographically on the properties of the dual SCFTs. We discuss a well known ``RG-monotone quantity" which can be computed both from the five-dimensional and higher-dimensional perspectives and shown to agree (as anticipated). Additionally, we comment on the higher-dimensional perspective on the computation of strip entanglement entropies and holographic Wilson loops. In particular, we emphasise the way in which the kinematical data of the UV CFT (the ranks of colour and flavour groups) ``decouples" from the details of the flow, at least for the particular quantities we study here.
It is in this sense that we are uncovering universal qualities of flows away from a large class of SCFTs.

We conclude in section \ref{concl} with some additional comments and a discussion of future directions this work will allow one to pursue.


\section{An  $\mathcal{N}=2$ Gauged SUGRA in $d=5$ from $\mathcal{N}=1$ Gauged SUGRA in $D=7$ }\label{sec:Nis2Theory}


In this section we start from the minimally supersymmetric $SU(2)$ gauged supergravity in seven dimensions \cite{Townsend:1983kk}, and perform a reduction to five dimensions on a two manifold $\Omega_{2,l}$. We write explicitly the reduced action as an $\mathcal{N}=2$ gauged supergravity in five dimensions, study its scalar potential and critical points.

\subsection{Reduction ansatz}

Our starting point is the minimal gauged supergravity (a theory with 16 supercharges) in seven dimensions \cite{Townsend:1983kk}, 
\begin{multline}
\mathcal{L} = R\hs 1-5 X^{-2} \dd X\wj\hs \dd X - \frac{1}{2}X^4 F_{(4)}\wj \hs F_{(4)}-\mathcal{V}\hs 1\\
-\frac{1}{2} X^{-2} F^i_{(2)}\wj \hs F^i_{(2)} - h F_{(4)}\wj B_{(3)}+\frac{1}{2} F^i_{(2)}\wj F^i_{(2)}\wj B_{(3)}.
\end{multline}
The field content is a metric, a scalar $X$, $SU(2)$ gauge fields $A^i_{(1)}$ with field strength $F^i_{(2)} = \dd A^i_{(1)}-\frac{\tilde{g}}{2}\as^{ijk} A_{(1)}^j\wj A_{(1)}^k$, and a three-form potential $B_{(3)}$ with field strength $F_{(4)} = \dd B_{(3)}$. The scalar potential is given by
\begin{equation}
\mathcal{V} = 2h^2 X^{-8}-4\sqrt{2}\tilde{g} h X^{-3} - 2\tilde{g}^2 X^2.
\end{equation}
Henceforth, we work in units such that the gauge coupling $\tilde{g} = 1$. The equations of motion following from the action of the supergravity theory (see appendix \ref{sec:eomsA}) are supplemented by an odd-dimension self-duality constraint
\begin{equation}
X^4\hs F_{(4)} = -2 h B_{(3)}+\frac{1}{2} A^i_{(1)}\wj F^i_{(2)}+\frac{1}{12}\as_{ijk} A^i_{(1)}\wj A^j_{(1)}\wj A^k_{(1)}.
\end{equation}

We now present a Kaluza--Klein reduction ansatz such that the resulting theory is an $\mathcal{N}=2$ gauged supergravity in five dimensions. The reduction of the metric takes the form
\begin{equation}
\dd s_7^2  = e^{-4\phi}\dd s^2_5 + e^{6\phi}\dd s^2 (\Omega_{2,l}).\label{papa}
\end{equation}
Here $\Omega_{2,l}$ is a Riemann surface (or quotient thereof). It is useful to work in the obvious orthonormal frame, where $\dd s^2_7 = \eta_{MN}\ee^M\ee^N$, in which the siebenbein can be taken as $\ee^M = (e^{-2\phi}\bar{\ee}^m, e^{3\phi}\bar{\ee}^a)$ with $m = (t,x,y,z,r)$ and $a = (1,2)$. Accordingly, we have 
\begin{equation}
\dd s^2_5 = \bar{\eta}_{mn} \bar{\ee}^m\bar{\ee}^n,\;\;\;\;
 \dd s^2(\Omega_{2,l}) = \bar{\ee}^a\bar{\ee}^a.\nonumber
 \end{equation}
The parameter $l=\{0,\pm1\}$ is related to the curvature of the Riemann surface ($S^2, T^2, H^2$) by $\dd \bar{\omega}^{12} = \mathcal{R}^{12} = l \mathrm{vol}_2$ in our conventions.
 
The ansatz for the $SU(2)$ gauge fields introduces an $SO(2)$ doublet of charged scalars $\theta$, and an $SO(2)$ gauge field $\mathcal{A}$. Crucially, the ansatz includes a twisting in this sector, in which these $SO(2)$ indices are aligned with the ``$a$" indices of the frame on $\Omega_{2,l}$:
\begin{eqnarray}
A^{1}  = - \as^{ab}\theta^a \bar{\ee}^b ,\;\;\;\;\;A^{2}  = \theta^a \bar{\ee}^a,\;\;\; \;\;A^3  = \bar{\omega}^{12} + \mathcal{A}.\nonumber
\end{eqnarray}
In terms of the $SO(2)$ covariant derivative $D\theta^a = \dd \theta^a-\as^{ab}\mathcal{A}\,\theta^b$, the corresponding field strengths are given by
\begin{eqnarray}
F^1  = -\as^{ab} D\theta^a\wj \bar{\ee}^b,\;\;\;
F^2  = D\theta^a\wj \bar{\ee}^a,\;\;\;\;F^3  = \mathcal{F} + (l -\theta\cdot\theta)\mathrm{vol}_2,\nonumber
\end{eqnarray}
where $\mathcal{F} = \dd \mathcal{A}$. 
The ansatz for the three-form potential is
\begin{equation}
B_{(3)} = c_{(3)}+\chi_{(1)}\wj \mathrm{vol}_2 -\frac{1}{2h}\dd(\bar{\omega}^{12}\wj\mathcal{A}).
\end{equation}
The last term, evidently singular in the $h\to 0$ limit, is not required when the topological mass is vanishing. In the case $h=0$, the three-form obeys a standard second order equation of motion.

Finally, the seven-dimensional scalar $X$ and the metric mode $\phi$ are taken to depend only on the coordinates of $\dd s^2_5$.  Let us write the five-dimensional action produced by this reduction.

\subsection{The $\mathcal{N}=2$ Action and Gauging}

After some calculation (outlined in appendix \ref{sec:eomsA}), we find that the equations of motion of the reduced theory can be obtained from the action,
\begin{multline}\label{eq:N2axn}
S_5 = \int \bar{R} \hf 1 -3\Sigma^{-2} \dd \Sigma \wj \hf \dd \Sigma - 2 \dd \varphi \wj \hf \dd \varphi - e^{2\varphi}D \theta^a \wj \hf D \theta^a \\
-\frac{1}{2}e^{4\varphi}\Big(\mathscr{D}\xi-\as^{ab}\theta^a D\theta^b\Big)\wj \hf \Big(\mathscr{D}\xi-\as^{ab}\theta^a D\theta^b\Big)- V_5 \hf 1\\
-\frac{1}{2}\Sigma^{-2}H^0\wj\hf H^0-\frac{1}{2}\Sigma^4H^1\wj\hf H^1
-\frac{1}{2}A^1\wj H^0\wj H^0.
\end{multline}
We introduced the redefined scalars $\Sigma$ and $\varphi$ such that
\begin{equation}
\Sigma = X e^{-2\phi},  \qquad \mathrm{and} \qquad e^\varphi = \frac{1}{X}e^{-3\phi},
\end{equation}
as well as the Stueckelberg scalar $\xi$ whose covariant derivative is defined as
\begin{equation}
\mathscr{D}\xi= \dd \xi + l A^0 +2h A^1.
\end{equation}
The vectors $A^0$ and $A^1$ are related to those of the reduction ansatz as $A^{\tilde{I}} = (\mathcal{A},-\chi_{(1)})$, and $H^{\tilde{I}} = \dd A^{\tilde{I}}$. 

The Hodge star $\hf$ is taken with respect to $\dd s^2_5$, and the five-dimensional potential is given by
\begin{equation}\label{eq:V5}
V_5 = -2\Sigma^2-4\sqrt{2}e^{2\varphi}\Sigma^{-1} h + 2e^{4\varphi}\Sigma^{-4} h^2-2 l e^{2\varphi}\Sigma^2+\frac{1}{2}e^{4\varphi}\Sigma^2 (l-\theta\cdot\theta)^2.
\end{equation}

This action is outwardly of the form required by $\mathcal{N}=2$ supersymmetry, with gravity plus one vector and one hyper multiplet. In our conventions (aligned with \cite{Bergshoeff:2004kh})   this theory takes the form
\begin{multline}
S_{\mathcal{N}=2} = \int R\hf 1-\frac{1}{2}g_{xy}D\phi^x\wj\hf D\phi^y-\frac{1}{2}g_{UV}Dq^U\wj\hf Dq^V-V_s\hf 1\\
-\frac{1}{2}a_{\tilde{I}\tilde{J}}H^{\tilde{I}}\wj\hf H^{\tilde{j}}-\frac{1}{3\sqrt{3}}\mathcal{C}_{\tilde{I}\tilde{J}\tilde{K}} A^{\tilde{I}}\wj H^{\tilde{J}}\wj H^{\tilde{K}}.
\end{multline}
Here, $x$ enumerates the scalars in the vector multiplets, $g_{xy}$ is the metric on a very special real manifold, $U$ enumerates the hypermultiplet scalars which are coordinates on a quaternionic-K\"ahler manifold with metric $g_{UV}$, and  $\tilde{I} = 0,\ldots,n_v$ where $n_v$ is the number of vector multiplets.

The vector multiplet scalar, which we identify with $\Sigma$, parametrises the very special real manifold $SO(1,1)$, while the four hypermultiplet scalars $q^U = (\varphi,\xi,\theta^1,\theta^2)$ parametrise the quaternionic-K\"ahler manifold $SU(2,1)/U(2)$, which is the so-called ``universal hypermultiplet". Some relevant details about the geometry of these manifolds can be found in appendix \ref{sec:geoA}.

The gauging of this theory is clearly within the hypermultiplet sector alone, and with $Dq^U = \dd q^U +A^{\tilde{I}}k_{\tilde{I}}^U$ we read off the Killing vectors associated with the gauged isometries as
\begin{equation}
k_0  = \, \theta^1\partial_{\theta^2}-\theta^2\partial_{\theta^1}+l\partial_\xi,\;\;\;\;\quad k_1  = \, 2h\,\partial_\xi.
\end{equation}
It is simple to see that these vectors generate $SO(2)\times \mathbb{R}$. The relevant terms in the scalar potential are given by
\begin{equation}\label{eq:Vs}
V_s = 8\left(\frac{3}{16}h^{\tilde{I}}k^U_{\tilde{I}}g_{UV}k^V_{\tilde{J}}h^{\tilde{J}}+ \vec{P}_x\cdot\vec{P}^x -2 \vec{P}\cdot \vec{P} \right),
\end{equation}
where the vectors
\begin{equation}
\vec{P}\equiv \frac{1}{2}h^{\tilde{I}}\vec{P}_{\tilde{I}} \qquad \mathrm{and} \qquad \vec{P}_x \equiv \frac{1}{2}h^{\tilde{I}}_x\vec{P}_{\tilde{I}}
\end{equation}
are computed from the $SO(1,1)$ data and the moment maps
\begin{equation}\label{eq:vPI}
\vec{P}_{\tilde{I}} = \frac{1}{2}\vec{J}_U\,^V\nabla_V k_{\tilde{I}}^U,
\end{equation}
with $\vec{J}$ the triplet of $SU(2)$ complex structures on $SU(2,1)/U(2)$. In appendix \ref{sec:eomsA} we compute these quantities explicitly, and verify that they reproduce the reduced scalar potential (\ref{eq:V5}) via (\ref{eq:Vs}). We discuss the critical points of the potential below,  in section \ref{sec:CP}.

\subsubsection{Minimal gauged SUGRA in $d=5$ from $D=7$}

We note in passing that these results can further be used to show that any solution of {\it minimal} gauged supergravity in $d=5$ can also be uplifted to the minimal theory in $D=7$. Subsequently these solutions can be promoted to Massive IIA or $D=11$ by the various formulae in the literature \cite{Lu:1999bc,Passias:2015gya}. This is accomplished by first taking
\begin{equation}
h\ne0,\quad l= -1, \quad \Sigma = (2h^2)^{1/6}, \quad \varphi = -\frac{1}{2}\ln\frac{3}{2}, \quad \theta^a = \xi =0 \quad \mathrm{and} \quad A^0 = 2h A^1.
\end{equation}
Introducing the redefined gauge field $\mathscr{A}$ with two-form field strength $\mathscr{H} = \dd\mathscr{A}$ such that
\begin{equation}
\mathscr{A} = A^1\sqrt{3}(2h^2)^{1/3}
\end{equation}
one finds that the resulting equations of motion can be derived from the action for minimal $\mathcal{N}=2$ gauged supergravity:
\begin{equation}
S_5 = \int\left(  \bar{R} +\frac{12}{L^2}\right) \hf 1
-\frac{1}{2}\mathscr{H}\wj\hf \mathscr{H}-\frac{1}{3\sqrt{3}}\mathscr{A}\wj \mathscr{H}\wj \mathscr{H}.
\end{equation}
This is a realisation of the conjecture from Gauntlett and Varela articulated in \cite{Gauntlett:2007ma}, see also \cite{Gauntlett:2006ai}. For a recently proposed proof, see \cite{Cassani:2019vcl}. This conjecture associates to each supersymmetric $AdS$ solution of $D=10$ or $D=11$ supergravity of the warped product form $AdS_{d+1}\times_w \mathcal{M}$ a truncation on $\mathcal{M}$ to the modes dual to the stress tensor multiplet in the corresponding SCFT$_d$. 

\subsection{Critical Points of the Scalar Potential}\label{sec:CP}

The scalar potential of the reduced theory has two critical points when $l=-1$, which corresponds to reducing on $H^2$. They both lead to $AdS_5$ geometries. The first is given by
\begin{equation}
\Sigma = (2h^2)^{1/6}, \quad \varphi = -\frac{1}{2}\ln\frac{3}{2}, \quad \theta^a = 0 \qquad \mathrm{such\,\,that} \qquad L^2_{AdS_5} = \frac{9}{2}(2h^2)^{-1/3}.\label{bpsx}
\end{equation}
We will see shortly that this solution is in fact supersymmetric. The second, which is not supersymmetric, is given by
\begin{align}
\Sigma & =\, \left(\frac{2h^2}{9} \right)^{1/6}\left [ -2 +(27\sqrt{17}+109)^{1/3}-(27\sqrt{17}-109)^{1/3}\right ]^{1/3},\nonumber\\
\varphi & = \frac{1}{2}\ln\left[\frac{2}{3}\left(-3+(\sqrt{17}+9)^{1/3}+(9-\sqrt{17})^{1/3} \right) \right],\nonumber\\
\theta^a & = 0.\label{nonbpsx}
\end{align}
This solution has previously appeared in \cite{Gauntlett:2002rv}\footnote{In fact, see around eq.(3.11) of the paper \cite{Gauntlett:2002rv}. We thank Jerome Gauntlett for pointing this out. }. We will briefly elaborate on the Physics associated to these exact solutions in section \ref{sec:QFT}. Let us now study the flows away from the supersymmetric fixed point.


\section{Supersymmetric Flows preserving four supercharges}\label{sec:Nis1flos}


Using the results of \cite{Ceresole:2001wi}, we next proceed to hunt for more general supersymmetric solutions in this theory. In particular, we will be interested in identifying domain wall type geometries which correspond holographically to renormalisation group flows preserving Poincar\'e invariance in 3+1 dimensions. 

In our conventions, the superpotential is given by
\begin{align}
W & = \sqrt{\frac{4}{3}\vec{P}\cdot \vec{P}}\nonumber\\
 & = \frac{1}{6}\Sigma^{-2}\sqrt{16e^{2\varphi}\,\theta\cdot\theta\, \Sigma^6+\Big(2\sqrt{2}\Sigma^3 + e^{2\varphi} \big(2h + \sqrt{2}(l-\theta\cdot\theta)\Sigma^3\big)\Big)^2}
\end{align}
Under certain conditions, including supersymmetry, one can derive the potential of the gauged supergravity via
\begin{equation}
V_s = -12 W^2 + 9 g^{\Omega \Xi}\partial_\Omega W \partial_\Xi W, 
\end{equation}
where $\Omega, \Xi$ label {\it all} scalars of the theory, i.e.  $\phi^\Omega = (\phi^x, q^U)$. This will be the case whenever $\partial_x\vec{Q} = 0$, which we define presently. 

The algebraic constraints for SUSY flows are written in terms of a vector $\vec{Q}$, which is a unit vector pointing along $\vec{P}$:
\begin{equation}
\vec{Q} \equiv \frac{\vec{P}}{|\vec{P}|} = \frac{2}{\sqrt{3}}\frac{\vec{P}}{W}.
\end{equation}
The full set of requirements for a solution to preserve at least four supersymmetries  is then given in terms of 
an ``RG flow" metric ansatz,
\begin{equation}
\dd s^2_5 = e^{2A}\left(-\dd t^2 +\dd \vec{x}^2 \right)+\dd r^2,
\end{equation}
as
\begin{align}
\partial_r A & = W,\\
\partial_x \vec{Q} & = 0, \label{eq:dQ}\\
\partial_r\phi^\Omega & = -3g^{\Omega\Xi}\partial_\Xi W.
\end{align}

The algebraic conditions from (\ref{eq:dQ}) are generically satisfied when $\theta^a = 0$. Note that when the potential can be written in terms of a superpotential, critical points of the superpotential are critical points of the potential--but the converse is not necessarily true. In particular, the second $AdS_5$ in (\ref{nonbpsx}) is not a critical point of the superpotential, and hence not supersymmetric. 

The flow equations for supersymmetric solutions in this reduced theory thus simplify to
\begin{eqnarray}
\partial_r A & = &\frac{\sqrt{2}}{3}\Sigma+\frac{1}{6}e^{2\varphi}\Big(l\sqrt{2}\Sigma+2h\Sigma^{-2}\Big),\label{eqAmain}\\
\partial_r \Sigma & = &-\frac{\sqrt{2}}{3}\Sigma^2 + \frac{1}{6}e^{2\varphi}\Big(4h\Sigma^{-1}-l\sqrt{2}\Sigma^2 \Big),\\
\partial_r\varphi & = &-\frac{1}{2}e^{2\varphi}\Big(l\sqrt{2}\Sigma+2h\Sigma^{-2} \Big).\label{eqvarphimain}
\end{eqnarray}
We now attempt to solve these equations.

\subsection{Analytic Solutions}

When $l=0$, corresponding to $\Omega_{2,0}\sim T^2$, the BPS equations can be solved exactly. Introducing a new radial coordinate $\rho$ such that
\begin{equation}
\frac{\partial\rho}{\partial r} = \Sigma^{-2},
\end{equation}
the BPS equations are readily integrated, yielding a general solution of the form
\begin{eqnarray}
& & \varphi  = -\frac{1}{2}\ln\left(2h\rho-c_\varphi\right),\;\;\;\;\Sigma  = \left(\frac{6 h\rho-3 c_\varphi }{3\sqrt{2}h\rho^2-3\sqrt{2}\rho c_\varphi+2 c_\Sigma}\right)^{1/3},\nonumber\\
& & A  = c_A +\frac{1}{6}\Big(\ln(2h\rho-c_\varphi)+2\ln\left(3\sqrt{2}h\rho^2-3\sqrt{2}\rho c_\varphi+2c_\Sigma \right) \Big),\label{estacv}
\end{eqnarray}
with the three integration constants $c_A, c_\varphi, c_\Sigma$. 

As a special case, when these integration constants vanish, one finds that the seven-dimensional scalars are given by
\begin{equation}
 X = \left(e^{-2\varphi}\Sigma^3 \right)^{1/5} = \left(8h^2\right)^{1/10},\;\;\;\; e^{-5\phi}=\Sigma e^{\varphi}=\left( 4^{1/5} h \rho \right)^{-5/6}.
\end{equation}
Furthermore, using our reduction formulae (\ref{papa}) to uplift this solution (with  $c_A=c_\varphi=c_\Sigma=0 $ and $h= \frac{1}{2\sqrt{2}}$) to seven dimensions, one finds a metric
\begin{equation}
\dd s^2_7 = \frac{3^{2/3}}{2^{5/6}}\rho\left(-\dd t^2 +\dd \vec{x}^2 \right) + \frac{1}{\sqrt{2}}\rho\,\dd s^2_{T^2}+\frac{2}{\rho^2}\dd \rho^2.\label{mana}
\end{equation}
This is in fact the supersymmetric $AdS_7$, with radius $L_7 = 2\sqrt{2}$.

For $l = \pm 1$, we have been unable to find closed-form solutions to the BPS equations. However, we have succeeded in reducing the full set of equations to a single Abel equation.
In fact, as explained in Appendix \ref{eqs5dappendix}, the whole system can be expressed in terms of the `radial variable' $\Sigma$. We have a configuration
\begin{eqnarray}
& & \dd s^2= e^{2A(\Sigma)}\dd x_{1,3}^2 + y^2(\Sigma){\dd\Sigma^2}\label{esta2main},\\
& & e^{2\varphi(\Sigma)}= \frac{1- y(\Sigma) f_1(\Sigma)}{y(\Sigma) f_2(\Sigma)}, \;\;\; \quad A(\Sigma)= \int \left[f_4(\Sigma) + e^{2\varphi} f_5(\Sigma)\right] y(\Sigma) \dd\Sigma.\nonumber
\end{eqnarray}
The functions $f_i(\Sigma)$ are defined in Appendix \ref{eqs5dappendix}, see eq.(\ref{fs}). The function $y(\Sigma)$ solves the Abel equation of the first kind,
\begin{eqnarray}
& & \partial_\Sigma y=  G_1 y + G_2 y^2 + G_3 y^3,\label{Abel}
\end{eqnarray}
where the functions $G_i(\Sigma)$ are defined in terms of the $f_i(\Sigma)$ in Appendix \ref{eqs5dappendix}, see eq.(\ref{Gs}). In this language, the solution in eqs.(\ref{estacv})-(\ref{mana}) corresponds (for vanishing integration constants) to,
\begin{eqnarray}
& & y= -\frac{3\sqrt{2}}{\Sigma^2},\;\;\;e^{2\varphi}=\Sigma^3,\;\; e^{2A}=\frac{1}{\Sigma^5},\;\;\; X=1,\;\;\; e^{-2\phi}=\Sigma.\nonumber\\
& & \dd s_7^2= \frac{1}{\Sigma^3}\left(\dd x_{1,3}^2+ \dd\Omega_{2,0}\right)+18\frac{\dd\Sigma^2}{\Sigma^2}\equiv e^{-\frac{v}{\sqrt{2}}} \left(\dd x_{1,3}^2+ \dd\Omega_{2,0}\right)+ \dd v^2.\nonumber
\end{eqnarray}
Let us now study other solutions.

\subsection{Numerical Solutions}

Absent an analytical solution, one can appeal to numerics to try to gain control over the landscape of supersymmetric flows. As an illustration, we focus here on flows with $l = -1$, in which the supersymmetric $AdS_5$ can provide a putative endpoint for the flow. For simplicity, in this section we choose $h=27/4$ such that the $AdS_5$ radius is one. To investigate the possibility for such flows, we note that the BPS equations, linearised around the SUSY $AdS_5$ vacuum, permit modes
\begin{equation}
A = r\left(1+\delta A\, r^{\hat{\delta}} \right),\qquad \Sigma = \Sigma_0+\delta\Sigma \,e^{r\hat{\delta}},\qquad \varphi = \varphi_0 + \delta\varphi \,e^{r\hat{\delta}},
\end{equation}
with 
\begin{equation}
\hat{\delta} = 0, -1\mp\sqrt{7}.
\end{equation}
The mode with $\hat{\delta} = 0$ is a universal mode that simply corresponds to a diffeomorphism resulting in an overall rescaling of all coordinates by the same factor. The modes $\hat{\delta} = -1\mp\sqrt{7}$ correspond to deformations of the $AdS_5$ that are compatible with supersymmetry. The upper sign is a relevant deformation, which would be naturally interpreted as a deformation of the dual $d=4$ SCFT, driving a flow towards the IR. The lower sign is an irrelevant deformation, and thus has a chance to describe the IR behaviour of a putative twisted $AdS_7\to AdS_5$ flow.

\subsubsection{AdS$_7\to$ AdS$_5$}

\begin{figure}
\centering
\includegraphics[scale=0.47]{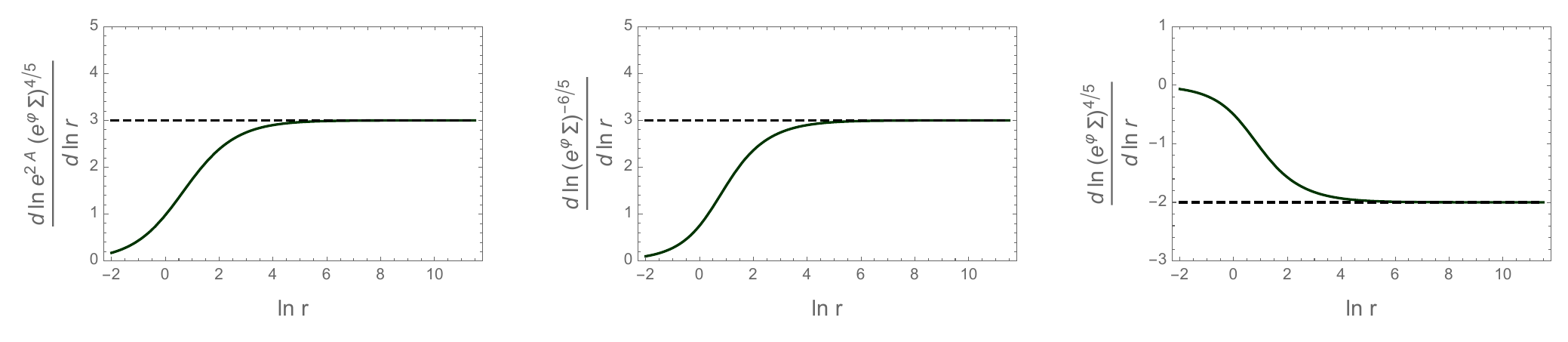}
\caption{\label{fig:725scale} UV scaling behaviour of various metric functions in the $D=7$ uplift of the solution to the $d=5$ BPS equations described in the text.}
\end{figure}

A flow realising the latter scenario has in fact already appeared in the literature \cite{Bah:2012dg}. Here we recover this solution from the perspective of our five-dimensional theory. By integrating the irrelevant mode out from the supersymmetric $AdS_5$ one obtains a seven-dimensional UV solution of the form
\begin{equation}
\dd s^2_7 \sim \frac{r^3}{\mathscr{L}^3} \Big(-\dd t^2 + \dd \vec{x}^2 + \dd s^2_{H^2} \Big) + \mathscr{L}^2\frac{\dd r^2}{r^2},
\end{equation}
for constant $\mathscr{L}$. This scaling behaviour is illustrated in figure \ref{fig:725scale}. Numerically, one finds that $\mathscr{L}^2 = \frac{9}{4}L^2_{7}$, with $L_{7}$ the supersymmetric $AdS_7$ radius.  Accordingly one can change radial variable such that $r = \rho^{2/3}$ and 
\begin{equation}
\dd s^2_7 \sim \frac{\rho^2}{\mathscr{L}^3} \Big(-\dd t^2 + \dd \vec{x}^2 + \dd s^2_{H^2} \Big) + L_{7}^2\frac{\dd \rho^2}{\rho^2},
\end{equation}
which is manifestly $AdS_7$ with two spatial directions wrapped on $H^2$. We thus have a numerical solution holographically describing the RG flow induced by wrapping a six-dimensional SCFT on $H^2$, which subsequently flows to a four-dimensional SCFT.

\subsubsection{AdS$_5\to?$}\label{sec:AdS5flo}

\begin{figure}[h]
\centering
\includegraphics[scale=0.29]{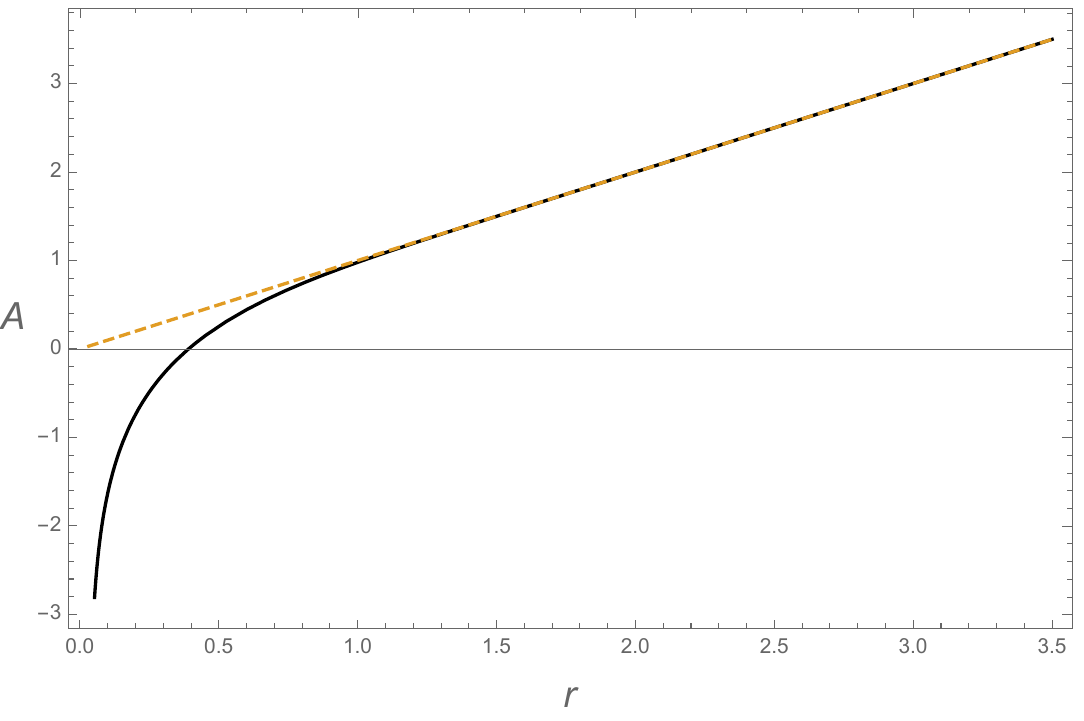}
\includegraphics[scale=0.29]{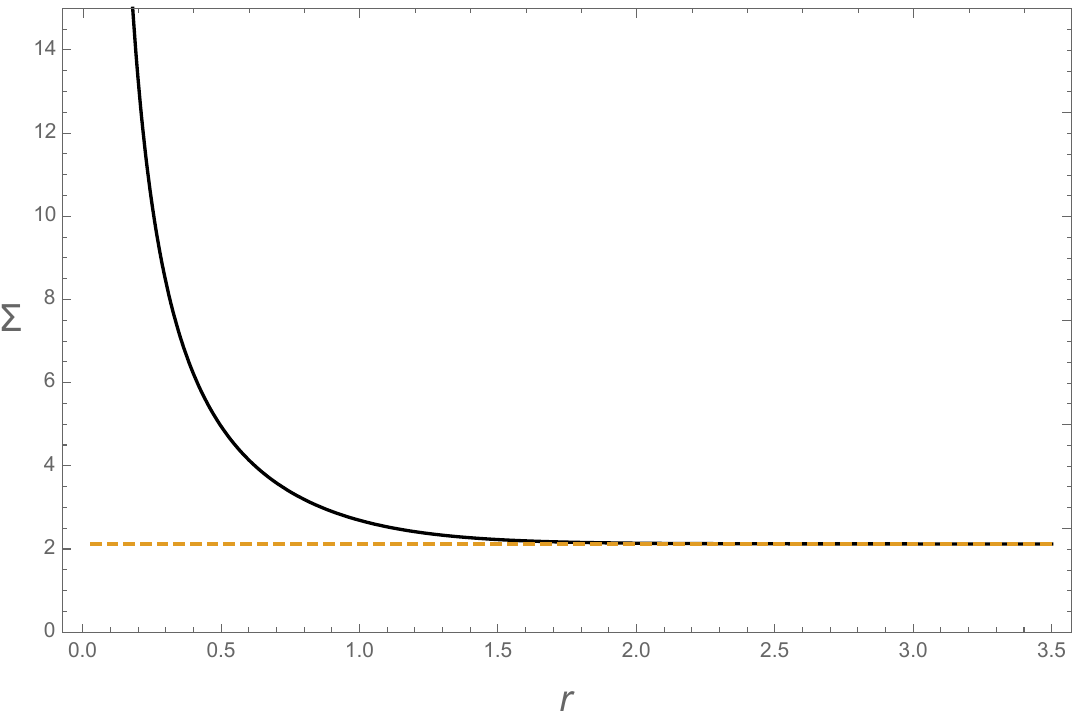}
\includegraphics[scale=0.29]{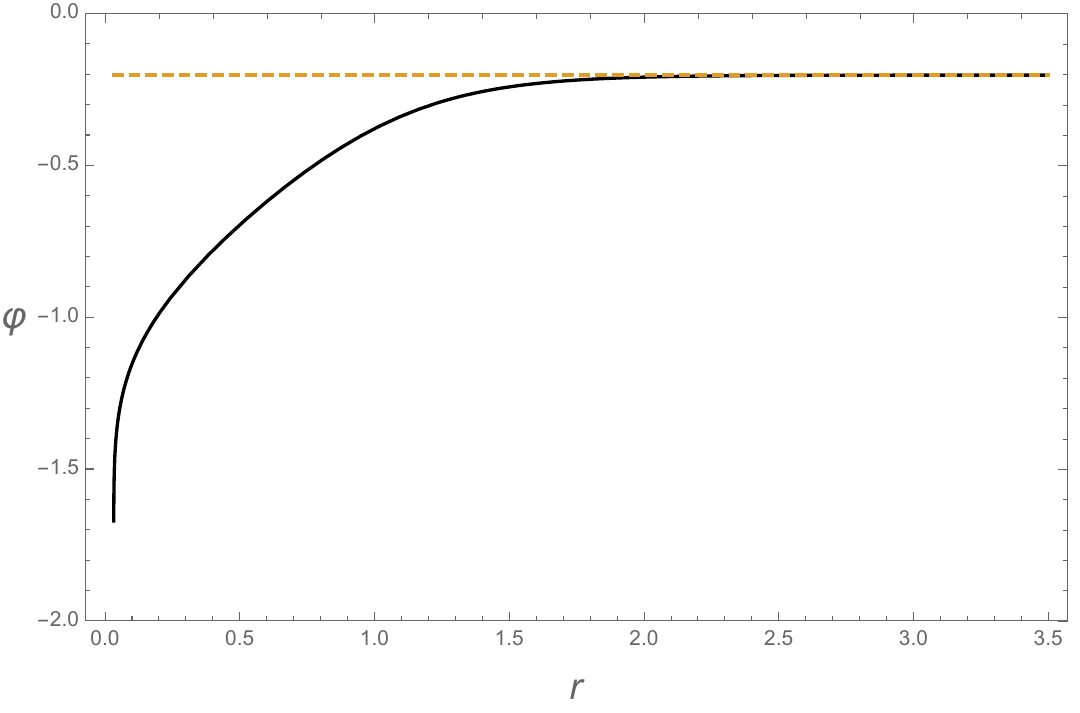}
\caption{\label{fig:susy5} The profiles of the bulk fields present in a supersymmetric flow from $AdS_5$ to the singular IR. The numerical solution is drawn in black, and the orange dashed lines give the asymptotic behavior of the fields at the $AdS_5$ critical point. The UV is approached as $r\to \infty$.}
\end{figure}
It is similarly straightforward to integrate the relevant mode, corresponding to the scalar operator of the dual SCFT$_4$ with dimension $\Delta = 1+\sqrt{7}$ acquiring a vev. The bulk solution is shown in figure \ref{fig:susy5}. The flow is singular, generically terminating at a fixed radial value. This IR singularity can be seen to be ``good" by the criterion in \cite{Maldacena:2000mw} once the solution is uplifted to eleven dimensions. Numerical evidence for this is given in figure \ref{fig:susy5IR} below--we outline the uplift  in section \ref{sec:10n11dUp}.

It is interesting to gain analytical control of the IR of this flow, which can give  important insights on its dual physics. From the numerical solution, we observe that as the IR is approached, $\Sigma \to \infty$ while $e^{2\varphi} \to 0$. In this limit, the leading form of the BPS equations (\ref{eqAmain})-(\ref{eqvarphimain}) is
\begin{align}
\partial_r A &= \frac{\sqrt{2}}{3}\left(1-\frac{1}{2}e^{2\varphi}\right)\Sigma,\\
\partial_r \Sigma & =-\frac{\sqrt{2}}{3} \left(1 - \frac{1}{2}e^{2\varphi}\right)\Sigma^2, \\
\partial_r\varphi & = \frac{1}{2}e^{2\varphi}\sqrt{2}\Sigma.
\end{align}
These coincide with those of the $h=0$ theory when $\theta = 0$ and $l = -1$ (see appendix \ref{sec:his0}). These equations can be solved exactly.  Indeed, using the new radial variable $\rho$ defined via $\partial_r\rho = -\Sigma$, the solution is
\begin{equation}
\varphi  = -\frac{1}{2}\ln\sqrt{2}\rho,\;\;\;\;
\Sigma  = s_0\frac{1}{\rho^{1/6}}e^{\frac{\sqrt{2}}{3}\rho},\;\;\;\;
A  = -\ln \Sigma+a_0.\label{solexcxx}
\end{equation}
The metric then approaches the form\footnote{Through an elementary change of radial coordinate, this metric can be thought of as a logarithmic correction to the IR of the flow in \cite{Pilch:2000ue} as given by \eqref{eq:CB}, which is dual to a state on the Coulomb branch of $\mathcal{N}=4$ SYM. We will find another example of this in section \ref{sec:8Qflo}.}
\begin{equation}
\dd s^2_{\mathrm{IR}} \sim \Sigma^{-2}\left(-\dd t^2+\dd \vec{x}^2 + \dd \rho^2 \right).
\end{equation}
In these coordinates, the singularity corresponding to the IR of the full flow is at $\rho\to\infty$. The coordinate change can be evaluated in the far IR limit to obtain
\begin{equation}
r-\mu \sim \frac{3}{\sqrt{2}}\rho^{1/6}e^{-\frac{\sqrt{2}}{3}\rho} = \frac{3}{\sqrt{2}}\frac{1}{\Sigma},
\end{equation}
in excellent agreement with the numerics\footnote{In the variables of eq.(\ref{esta2main}) the solution is harder to find as the resolution of the Abel equation (\ref{Abel}) leads to a transcendental equation for $y(\Sigma)$.}. 

We can analyse the character of the singularity using the criterium that Gubser presented in
\cite{Gubser:2000nd}. A ``good", or acceptable singularity requires that the potential in eq.(\ref{eq:V5}) is bounded above for the asymptotics in eq.(\ref{solexcxx}).
Considering eq.(\ref{solexcxx}), near the singularity the leading order behaviour of the potential is given by
\begin{equation}
V= \Sigma^2\left( -2 -2 l e^{2\varphi} +\frac{e^{4\varphi}}{2} l^2 -4 h \sqrt{2} \frac{e^{2\varphi}}{\Sigma^3} +2h^2\frac{e^{4\varphi}}{\Sigma^6}\right)\sim -2\Sigma^2\to-\infty.
\end{equation}
The potential is bounded above, and hence the singularity acceptable according to Gubser's criterium.

In summary, we can view this supersymmetric solution as an RG flow which connects the supersymmetric $AdS_5$ in the UV to an exact solution of the $h=0$ theory in the IR. We shall study some physical aspects of this solution in section \ref{sec:QFT}. Now, let us study the lift of these backgrounds to eleven or ten-dimensional supergravities.


\section{Uplifts to 10 and 11 dimensions}\label{sec:10n11dUp}


In this section we describe the lift of the flows described by the different solutions of eqs. (\ref{eqAmain})-(\ref{eqvarphimain}). We first describe the lift to eleven-dimensional supergravity. After that, we discuss the lift to Massive IIA supergravity. As previously advertised, there are actually `infinitely many' lifts to Massive IIA. We shall clarify this statement below.

\subsection{The $\mathcal{N}=2$ gauged supergravity uplift to $D=11$}

For non-vanishing topological mass $h$, the uplift of the minimal gauged supergravity in $D=7$ to $D=11$ was given in \cite{Lu:1999bc}. To align their conventions with ours, we first take $\tilde{g}\to g$ and $h\to\frac{1}{2\sqrt{2}}g$. We continue to work in units where $\tilde{g}=g=1$. It will also be convenient to recall that the seven-dimensional scalar $X$ is related to those of the five-dimensional theory like
\begin{equation}
X = \left(e^{-2\varphi}\Sigma^3 \right)^{1/5},\;\;\;\; e^{-5\phi}= \Sigma e^{\varphi}.\label{japij}
\end{equation}
We also remind the reader that the seven-dimensional metric is given by
\begin{equation}
\dd s^2_7 = \left(e^\varphi\Sigma \right)^{4/5}\dd s^2_5 + \left(e^\varphi\Sigma \right)^{-6/5}\dd \Omega^2_{2,l}.\label{panana}
\end{equation}
Let us move now to the details of the lift. We  introduce an interval coordinate $\zeta$, and define
\begin{equation}
\Delta = X^{-4}\sin^2\zeta+X \cos^2\zeta.\label{www12}
\end{equation}
The internal manifold in the uplift of the seven-dimensional metric is then given by
\begin{equation}
\dd s^2_4 = X^3 \Delta \dd\zeta^2+\frac{1}{4}X^{-1}\cos^2\zeta\, \vec{\Omega}\cdot\vec{\Omega},
\end{equation}
where 
\begin{equation}
\vec{\Omega} =\vec{\sigma}-\vec{A}_{(1)},
\end{equation}
with the $\vec{\sigma}$ left-invariant one-forms on $S^3$ such that $\dd\sigma^i = -\frac{1}{2}\as_{ijk}\sigma^j\wj\sigma^k$ and the $\vec{A}_{(1)}$ are the $SU(2)$ field strengths of the $D=7$ theory.
With these considerations, the $D=11$ metric is
\begin{equation}
\dd s^2_{11} = \Delta^{1/3}\dd s^2_7 + 2\Delta^{-2/3}\dd s^2_4,\label{www34}
\end{equation}
and the three-form flux can be written as
\begin{multline}
\hat{A}_{(3)} = -\sin\zeta\Big(\left(\mathscr{D}\xi -\as^{ab}\theta^aD\theta^b\right)\wj\mathrm{vol}_2+\Sigma^4\hf \dd\chi_{(1)}\Big)\\-\frac{1}{\sqrt{2}}\sin\zeta\Big(-\as^{ab}D\theta^a\wj\bar{\ee}^b\wj\Omega^1 +D\theta^a\wj\bar{\ee}^a\wj\Omega^2+\big(\mathcal{F}+(l-\theta\cdot\theta)\mathrm{vol}_2\big)\wj\Omega^3\Big)\\
+\frac{1}{2\sqrt{2}}\Big(2\sin\zeta+\sin\zeta\cos^2\zeta\Delta^{-1} X^{-4}\Big)\Omega^1\wj\Omega^2\wj\Omega^3.
\end{multline}

\begin{figure}
\centering
\includegraphics[scale=0.35]{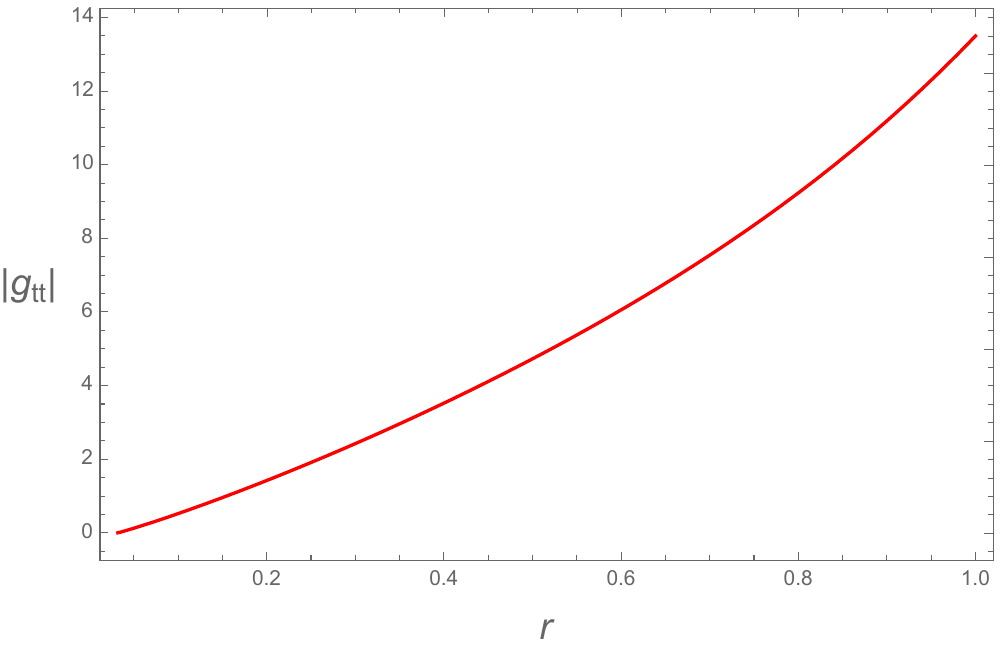}
\caption{\label{fig:susy5IR} The IR behaviour of $|g_{(11)}\,_{tt}|$, as given by \eqref{eq:gtt11susy5}, for the supersymmetric flow once lifted to eleven dimensions. That this component of the metric vanishes at the singularity is consistent with the singularity being ``good" in the classification of \cite{Maldacena:2000mw}. Pictured is the $\zeta = \pi/6$ slice.}
\end{figure}

Note that for any uplifted five-dimensional solution, the singularity criterion of \cite{Maldacena:2000mw} requires that 
\begin{equation}\label{eq:gtt11susy5}
|g_{(11)}\,_{tt}| = e^{\frac{2}{3}\varphi}\Sigma\,|g_{(5)}\,_{tt}|\left(\cos^2\zeta+e^{2\varphi}\Sigma^{-3}\sin^2\zeta \right)^{1/3}
\end{equation}
does not increase anywhere along $\zeta$ as the far IR is approached. An example of its behaviour is shown in figure \ref{fig:susy5IR}.

Let us discuss now the lift to Massive IIA.

\subsection{The $\mathcal{N}=2$ gauged supergravity uplift to $D=10$}\label{liftmassiveiia}

Any solution with non-vanishing $h$ can also be uplifted via the formulae in \cite{Passias:2015gya}  to the Massive IIA Supergravity theory in ten dimensions. In fact, this can be accomplished in infinitely many ways (all possible flavour and colour groups combinations defining a six-dimensional ${\cal N}=(1,0)$ SCFT). As in the eleven-dimensional uplift, we continue to work in units with $g=1$ and $h=\frac{1}{2\sqrt{2}}$. For brevity, we quote below only the metric and dilaton in Massive IIA, using the variables in \cite{Cremonesi:2015bld}. In Appendix \ref{appendixmassiveiia} we give a full account of the configuration.

Recall that the configuration in seven dimensions is described by the fields $\Sigma(r), \varphi(r), A(r)$ satisfying BPS equation (\ref{eqAmain})-(\ref{eqvarphimain}). It was useful to define the combinations  $X^5$ and $e^{-5\phi}$ as in eq.(\ref{japij}). The seven-dimensional metric  is of the form given in eq.(\ref{panana}).
%
The gauge field $A^{3}= \bar{\omega}^{12}$ is the spin connection on $\dd s^2 (\Omega_{2,l})$. The metric and dilaton (and all other fields) in Massive IIA are  given in terms of the functions $X(r)$ and  $\alpha(z)$ (together with its derivatives $\dot{\alpha}$ and $\ddot{\alpha}$). The function $\alpha(z)$ is defined in an interval $0\leq z\leq z_*$ and encodes all the field-theoretical information that defines a minimally SUSY six-dimensional CFT, see \cite{Cremonesi:2015bld,Filippas:2019puw} for the details. The metric and dilaton in Massive IIA read
\begin{eqnarray}
& & \frac{1}{\mu}\dd s_{st}^2=  X^{-1/2} \sqrt{-\frac{\alpha}{\ddot{\alpha}}} \dd s_7^2 + X^{5/2} \sqrt{-\frac{\ddot{\alpha}}{{\alpha}}}\left(dz^2+\left( \frac{\alpha^2}{\dot{\alpha}^2- 2 \alpha \ddot{\alpha}}\right)\frac{1}{W}   D\Omega_2\right),\nonumber\\
& & e^{2\Phi}= e^{2\Phi_0} X^{5/2} \left( -\frac{\alpha}{\ddot{\alpha}}\right)^{3/2} \frac{1}{W (\dot{\alpha}^2-2\alpha\ddot{\alpha})},\label{lift10xxx}\\
& & W=(1-X^5)\frac{\dot{\alpha}^2}{ (\dot{\alpha}^2-2\alpha\ddot{\alpha})       } + X^5,\;\;\; D\Omega_2= d\chi^2 +\sin^2\chi(d\psi- \frac{1}{2}A^{3})^2.\nonumber
\end{eqnarray}
The numbers $\mu,\Phi_0$ can be read from eqs. (2.3) and (2.4) in \cite{Filippas:2019puw}. Notice that when $X=1$ we have $W=1$ and this solution with $A^{3}=0$ is the uplift of the configuration in eq.(\ref{estacv}), that is $AdS_7\times M_3$. The function $\alpha(z)$ is then determined by all the possible choices of six-dimensional SCFT. It is in this sense that we have infinitely many lifts to Massive IIA.
\\
The uplifted solution (if singular) will present an acceptable singularity if
\begin{equation}
g_{tt,E}= e^{-\Phi_0/2} \left(-\frac{\alpha}{\ddot{\alpha}} \right)^{1/8} (\dot{\alpha}^2-2\alpha\ddot{\alpha})^{1/4} W^{1/4} X^{-9/8} e^{4\varphi/5}\Sigma^{4/5} e^{2A},
\end{equation}
is bounded close to the singular point, for any value of $z$.

Other details of this uplift, such as the expression for the $B_2$ and the Ramond fields, are given in Appendix \ref{appendixmassiveiia}. 
 We now turn to a different five-dimensional supergravity and its uplift.


\section{An $\mathcal{N}=4$ Gauged SUGRA in $d=5$ from $\mathcal{N}=2$ Gauged SUGRA in $D=7$}\label{sec:Nis4}


In \cite{Cheung:2019pge}, \cite{Cassani:2019vcl}  a consistent truncation of eleven-dimensional supergravity to a five-dimensional supergravity theory with $\mathcal{N}=4$ supersymmetry was presented. As in the previous section, this theory can be obtained by reducing a seven-dimensional gauged supergravity on a Riemann surface with a topological twist incorporated into the reduction ansatz. In this case, however, the starting point is the $SO(5)$ gauged supergravity in seven dimensions, which can be uplifted to eleven dimensions according to \cite{Nastase:1999cb,Nastase:1999kf}. 

Here too the five-dimensional theory can be efficiently used to identify $\frac{1}{2}$-BPS solutions holographically dual to RG flows in a $(3+1)$-dimensional field theory. In this section we discuss such RG flows.

\subsection{Setup}

The five-dimensional theory is an $\mathcal{N}=4$ gauged supergravity with three vector multiplets and non-compact gauge group $ SO(2)\times SE(3)$. In our conventions, the Einstein-scalar sector of the theory is described by the Lagrangian
\begin{equation}
\mathcal{L}_{\mathrm{scalar}} = R\hs1 -3\Sigma^{-2}\dd \Sigma\wj\hs \dd \Sigma + \frac{1}{8} D\mathcal{M}_{MN}\wj \hs D\mathcal{M}^{MN}+\mathcal{L}_{\mathrm{pot}},
\end{equation}
where
\begin{multline}
\mathcal{L}_{\mathrm{pot}} = \frac{1}{2} g^2\Bigg[f_{MNP}f_{QRS}\Sigma^{-2}\Big(\frac{1}{12}\mathcal{M}^{MQ}\mathcal{M}^{NR}\mathcal{M}^{PS}-\frac{1}{4}\mathcal{M}^{MQ}\eta^{NR}\eta^{PS} +\frac{1}{6}\eta^{MQ}\eta^{NR}\eta^{PS}\Big)\\
+\frac{1}{4}\xi_{MN}\xi_{PQ}\Sigma^4\Big(\mathcal{M}^{MP} \mathcal{M}^{NQ}-\eta^{MP}\eta^{NQ} \Big)+\frac{1}{3}\sqrt{2}f_{MNP}\xi_{QR} \Sigma \mathcal{M}^{MNPQR}\Bigg].
\end{multline}
The capital Roman indices from the middle of the alphabet are $SO(5,3)$ indices, and here we will raise and lower them with the metric $\eta = \mathrm{diag}(-,-,-,-,-,+,+,+)$. Capital indices from the beginning of the alphabet index $SO(5)\times SO(3)$ (in this order), and we will often split them such that e.g. $A \in \{m,a\}$ where $m=1,2,\ldots5$ and $a = 6,7,8$. The symmetric tensor $\mathcal{M}_{MN}$ can be constructed from a vielbein $\mathcal{V}^A\,_M$ parametrising the coset $SO(5,3)/(SO(5)\times SO(3))$ in the standard way,
\begin{equation}
\mathcal{M} = \mathcal{V}^T\mathcal{V},
\end{equation}
and we employ also the completely antisymmetric tensor $\mathcal{M}_{MNPQR} = \as_{m_1\ldots,m_5}\mathcal{V}^{m_1}\,_M\ldots\mathcal{V}^{m_5}\,_R$.

Given a parametrisation of the scalar manifold, the remaining data specifying the supergravity theory is contained within the embedding tensors, which determine the gauging. Here, 
\begin{equation}
\xi_{45} = -\sqrt{2}
\end{equation}
and
\begin{align}
f_{123} = -\frac{1}{2}(3+l), \quad f_{678} = \frac{1}{2}(3-l), \quad f_{128} =f_{236}=-f_{137}= -\frac{1}{2}(l+1),\nonumber\\
f_{178} = -f_{268}=f_{367}=\frac{1}{2}(1-l),
\end{align}
where again $l=\{0,\pm1\}$ is a twisting parameter in the reduction. These tensors are totally antisymmetric.

\subsection{Supersymmetric flows preserving eight supercharges}\label{sec:8Qflo}

The conditions for $\frac12$-BPS domain walls (preserving eight of the sixteen supercharges) are given in \cite{Bobev:2018sgr} in terms of the dressed embedding tensors, defined as
\begin{equation}
\hat{f}^{ABC}  = f^{MNP}\mathcal{V}^A\,_M \mathcal{V}^B\,_N \mathcal{V}^C\,_P, \qquad \mathrm{and}\qquad \hat{\xi}^{AB} = \xi^{MN}\mathcal{V}^A\,_M \mathcal{V}^B\,_N.
\end{equation}
Any such flow will satisfy the algebraic constraints
\begin{align}
P^{[mn}P^{pq]} & = 0,\\
\partial_\Sigma\left(W^{-1}P^{mn}\right) & = 0,\\
\hat{f}^{amn}P_{mn}& = 0,\\
\frac{1}{4\sqrt{2}}\Sigma^3\as^{mn}\,_{pqr}P^{pq}\hat{\xi}^{ra} & = P_p\,^{[m}\hat{f}^{n]pa}
\end{align}
and the BPS equations
\begin{align}\label{8QBPS1}
A' & = W,\\
\Sigma' & = -\Sigma^2 \partial_\Sigma W,\\
\phi^{x}\,' & = -3 g^{xy}\partial_y W.\label{8QBPS3}
\end{align}
In writing these, we have introduced the shift matrix and superpotential
\begin{equation}
P^{mn}  = -\frac{1}{6\sqrt{2}}\Sigma^2\hat{\xi}^{mn}+\frac{1}{36}\Sigma^{-1}\as^{mnpqr}\hat{f}_{pqr},\qquad W = \sqrt{2P^{mn}P_{mn}},
\end{equation}
again parametrised the domain wall solutions as
\begin{equation}
\dd s^2_5 = e^{2A}\left(-\dd t^2 +\dd \vec{x}^2 \right)+\dd r^2,
\end{equation}
and labeled the surviving scalars in the truncated $SO(5,3)/(SO(5)\times SO(3))$ coset $\phi^x$ with metric $g_{xy}$. For such solutions, in our conventions, the scalar potential $V_{s}$ can be written in terms of the superpotential as
\begin{equation}
V_{s} = 9g^{xy}\partial_x W\partial_y W + 3\Sigma^2 \left(\partial_\Sigma W\right)^2-12 W^2.
\end{equation}

It is straightforward to demonstrate that an $SO(2)$ invariant sector of the $\mathcal{N} = 4$ theory immediately satisfies all of the algebraic constraints.\footnote{There may be alternative ways to satisfy the necessary conditions. This however appears to be the largest subtruncation, based on invariant subgroups, that automatically satisfies all of the algebraic constraints.} This sector is a further consistent truncation of the $\mathcal{N}=4$ theory to the modes invariant under the $SO(2)\subset SO(2)\times SE(3)$, leaving a total of seven scalars.\footnote{We ignore the Stueckelberg scalars $\xi^\alpha$ of \cite{Cheung:2019pge} throughout, which will play no role in the RG flow solutions of immediate interest.} In this reduced theory, the metric on the scalar manifold is given by
\begin{equation}
\dd s^2_{SO(2)} = 3\Sigma^{-2}\dd \Sigma^2+3\dd\varphi_3^2 + \frac{1}{4}\mathcal{T}^{-1}_{\alpha\beta}\mathcal{T}^{-1}_{\gamma\delta}\dd\mathcal{T}_{\beta\gamma}\dd \mathcal{T}_{\delta\alpha},
\end{equation}
where the symmetric tensor $\mathcal{T}$  can be constructed as $\mathcal{T} = V^T V$ with
\begin{equation}
V=
\begin{pmatrix}
e^{\varphi_1} & e^{\varphi_1}a_1 & e^{\varphi_1}(a_1 a_2 + a_3) \\
0 & e^{\varphi_2-\varphi_1} & e^{\varphi_2-\varphi_1} a_2 \\
0 & 0 & e^{-\varphi_2}
\end{pmatrix}.
\end{equation}
The superpotential for the full $SO(2)$ invariant sector is given by
\begin{multline}
W = \frac{g}{6\Sigma}e^{-2(\varphi_1+\varphi_2)-3\varphi_3}\Big(e^{2(\varphi_1+\varphi_3)}+\left(1+a_1^2(1+a_2^2)+2 a_1 a_2 a_3+a_3^2 \right)e^{2(2\varphi_1+\varphi_2+\varphi_3)}\\+(1+a_2^2)e^{4\varphi_2+2\varphi_3}+le^{2(\varphi_1+\varphi_2)}+2\Sigma^3 e^{2\varphi_1+2\varphi_2+3\varphi_3} \Big)
\end{multline}
and the first order BPS equations are those dictated by \eqref{8QBPS1}-\eqref{8QBPS3}. From the structure of the superpotential it is easy to see that it is consistent to switch off the three axions $a_i=0$ and keep the three dilatons $\varphi_i$ plus the scalar in the gravity multiplet $\Sigma$. The remaining equations are thus 
\begin{align}
A' & =  \frac{g}{6\Sigma}e^{-2\varphi_2-\varphi_3}\Big(1+e^{2\varphi_1+2\varphi_2}+e^{-2\varphi_1+4\varphi_2}+le^{2\varphi_2-2\varphi_3}+2\Sigma^3 e^{2\varphi_2+\varphi_3} \Big),\\
\Sigma' & =  \frac{g}{6}e^{-2\varphi_2-\varphi_3}\Big(1 +e^{2\varphi_2+2\varphi_1}+ e^{-2\varphi_1+4\varphi_2}+le^{2\varphi_2-2\varphi_3}-4\Sigma^3 e^{2\varphi_2+\varphi_3} \Big),\\
\varphi_1' & = \frac{g}{3\Sigma}e^{-2\varphi_2-\varphi_3}\Big(1 + e^{-2\varphi_1+4\varphi_2}-2e^{2\varphi_1+2\varphi_2} \Big),\\
\varphi_2' & = \frac{g}{3\Sigma}e^{-2\varphi_2-\varphi_3}\Big(2 - e^{-2\varphi_1+4\varphi_2}-e^{2\varphi_1+2\varphi_2} \Big),\\
\varphi_3' & = \frac{g}{6\Sigma}e^{-2\varphi_2-\varphi_3}\Big(1 +e^{2\varphi_2+2\varphi_1}+ e^{-2\varphi_1+4\varphi_2}+3le^{2\varphi_2-2\varphi_3} \Big).
\end{align}
The $AdS_5$ preserving 16 supersymmetries is recovered by setting the twisting parameter $l = -1$, as well as $\varphi_i = 0$, $\Sigma =1/2^{1/3} $ and $A = r/L$ with $L$ the $AdS$ radius, $L = 2^{2/3}/g$. We are primarily interested in flows from this $AdS_5$ triggered by sources for relevant operators, or expectation values of operators in the dual field theory. We thus fix $l = -1$ from this point on, although analytic solutions for other values of $l$ can also be found in some simple cases. 

 A linearised analysis around the $AdS$ solution shows that the dilaton modes permitted by supersymmetry correspond to vevs for two $\Delta=4$ operators and a source for a $\Delta=6$ operator, while the scalar in the gravity multiplet $\Sigma$ is dual to the vev of a $\Delta=2$ operator.

The BPS equation can be solved in the radial coordinate $z$ defined as
\begin{equation}\label{eq:drdzNis4}
\frac{\dd r}{\dd z} = -\frac{2}{g}\frac{\Sigma}{z}e^{2\varphi_2+\varphi_3}
\end{equation}
Notice that the last three equations decouple from the rest. The most general solution in this sector contains three integration constants. If we want to keep $AdS_5$ either as the UV or IR of the flow, then it is clear that we cannot keep simultaneously the source for the irrelevant operator and the marginal vevs. Switching off the vevs we obtain a flow that reaches $AdS_7$ in the UV, first found in \cite{Maldacena:2000mw}.

If instead we keep only the vevs, we have $AdS_5$ as the UV geometry. The generic solution in this class will have a bad IR singularity. However, it is possible to tune the vevs, by choosing either of the integration constants for $\varphi_1$ or $\varphi_2$ to vanish, so that the IR behaviour is softened---still singular, but ``good" in the classification of \cite{Maldacena:2000mw}. This particular solution, once the irrelevant source is switched off in order to retain a CFT in the UV, reads
\begin{align}
e^{-6\varphi_2} = & \, \left(1+c_\varphi\, z^4 \right),\\[2mm]
e^{2\varphi_3} = & \, \frac{(1+c_\varphi \,z^4)^{1/6}\arcsinh\left(\sqrt{c_\varphi}\,z^2\right)}{\sqrt{c_\varphi}\,z^2},\\[2mm]
\Sigma = & \, \frac{\left(c_\varphi z^4\right)^{1/12}\left(1+c_\varphi\,z^4\right)^{1/12}\left[\arcsinh\left(\sqrt{c_\varphi}\,z^2\right)\right]^{1/6}}{\left[2z^4c_\sigma\sqrt{c_\varphi}+z^2\sqrt{c_\varphi}\left(1+c_\varphi\,z^4\right)^{1/2}+\arcsinh\left(\sqrt{c_\varphi}\,z^2\right)\right]^{1/3}},\\[2mm]
\frac{e^{6A}}{L^6} =&\frac{(1+c_\varphi \,z^4)^{1/2}\arcsinh\left(\sqrt{c_\varphi}\,z^2\right)\left[2z^4c_\sigma\sqrt{c_\varphi}+z^2\sqrt{c_\varphi}\left(1+c_\varphi\,z^4\right)^{1/2}+\arcsinh\left(\sqrt{c_\varphi}\,z^2\right)\right]}{2c_\varphi\,z^{10}}.
\end{align}
The remaining dilaton assumes the profile 
\begin{equation}
e^{-3\varphi_1}=\left(1+c_\varphi\, z^4 \right)\qquad\qquad{\rm or}\qquad\qquad e^{6\varphi_1}=\left(1+c_\varphi\, z^4 \right)
\end{equation}
depending on which integration constant for $\varphi_1$ or $\varphi_2$ one chooses to switch off. Notice that since we forced the irrelevant source to vanish and tuned one of the vevs there is only one free parameter $c_\varphi$ from the dilaton sector. 

It is straightforward to see that the UV of this solution is indeed the known supersymmetric $AdS_5$, while the (singular) IR is generically described by a metric of the form
\begin{equation}
\dd s_5^2\sim\left(\frac{\log z}{z^4}\right)^{1/3}\left(-\dd t^2+\dd\vec{x}^2+\frac{\log z}{z^6}\,\dd z^2\right).
\end{equation}
With a careful tuning of the vevs such that $2c_\sigma=-\sqrt{c_\varphi}$, however, the leading singularity can be removed. In this case, the singular IR geometry becomes
\begin{equation}\label{eq:logCB}
\dd s_5^2\sim\left(\frac{\log z}{z^4}\right)^{2/3}\left(-\dd t^2 +\dd\vec{x}^2+\frac{\dd z^2}{z^2}\right),
\end{equation}
which can be seen, in this radial coordinate, as a logarithmic correction to the Coulomb branch metric, discussed further around \eqref{eq:CB}.

In either case, these IR geometries do not exhibit (hyper)scaling. For the metric given in  (\ref{eq:logCB}), one can solve exactly the massless Klein-Gordon equation for a scalar mode in this background. Repeating the line of arguments that will be developed in section \ref{sec:spectral}, one thus anticipates that the solution with this IR behaviour holographically describes a gapped phase of the dual field theory.

\subsection{Uplift to 11D}

The uplift to eleven dimensions can be efficiently accomplished in a two step process. First, we lift to solutions of seven-dimensional gauged supergravity, then we uplift on $S^4$ \cite{Nastase:1999cb,Nastase:1999kf}. The seven-dimensional metric is given by
\begin{equation}
\dd s^2_7 = \left(\frac{\Sigma}{e^{3\varphi_3}}\right)^{2/5}\dd s^2_5 + \left(\frac{e^{3\varphi_3}}{\Sigma} \right)^{3/5}\dd s^2(H^2)
\end{equation}
and the surviving scalars enter the symmetric matrix $T_{ij}$, with index $i\in \{a,\alpha\}$ like
\begin{equation}
T_{ab} = \Big(e^{\varphi_3}\,\Sigma^3 \Big)^{3/5}\delta_{ab}, \quad T_{a\alpha} = 0, \quad T_{\alpha\beta} = \left(\frac{1}{e^{\varphi_3} \, \Sigma^3} \right)^{2/5}
\begin{pmatrix}
e^{2\varphi_1} & 0 & 0\\
0 & e^{2\varphi_2-2\varphi_1} & 0\\
0 & 0 & e^{-2\varphi_2}
\end{pmatrix}.
\end{equation}
The eleven-dimensional metric, then, is given by
\begin{equation}
\dd s^2_{11} = \Delta^{1/3}\dd s^2_7+\frac{1}{g^2}\Delta^{-2/3}T^{-1}_{ij}D\mu^i D\mu^j,
\end{equation}
where $\mu^i\mu^i = 1$ are constrained coordinates on $S^4$, and 
\begin{equation}
\Delta = T_{ij}\mu^i\mu^j, \qquad D\mu^a = \dd \mu^a + \bar{\omega}^{ab}\mu^b, \qquad D\mu^\alpha = \dd\mu^\alpha,
\end{equation}
with $\bar{\omega}^{ab}$ the spin connection on $H^2$. 

This is enough to check the nature of the IR singularity. We see that the temporal component of the metric vanishes as $\left(\log z\right)^{-1/3}$ for the general solution and $z^{-4/3}$ for the one with improved asymptotics (which is the same as the Coulomb branch flow, equation \eqref{eq:CB}). 

The uplift we have presented, with non-trivial dilatons $\varphi_i$, corresponds to RG flows in the $\mathcal{N}=2$ CFT of \cite{Maldacena:2000mw}. On the other hand, the solution with $c_\varphi = 0$ is supported solely by $\Sigma$, which is the scalar in the gravity multiplet. This scalar can be identified with the one present in Romans five-dimensional supergravity \cite{Romans}, so this particular solution can be obtained in that restricted theory. As a consequence it has two possible uplifts: either to type IIB supergravity using \cite{Lu:1999bw} (see \cite{Bobev:2019ylk} for explicit formulae), resulting in the $SO(4)\times SO(2)$ invariant Coulomb branch flow of \cite{Freedman:1999gk}, or to eleven-dimensional supergravity using the expressions in \cite{Gauntlett:2007sm}. According to this last reference, Romans supergravity is a consistent truncation on any manifold in the class found in \cite{LLM}, so the uplift of this solution corresponds to RG flows, triggered by the vev of a dimension 2 operator, on an arbitrary Gaiotto--Maldacena gauge theory. 

The eleven-dimensional metric in this restricted solution reads
\begin{equation}\label{eq:CBuplift}
\dd s_{11}^2=\frac{\widetilde{\Delta}^{1/3}}{\lambda\,\widetilde{X}^{1/3}}\left(\dd s_5^2+\dd s_6^2\right)\,,
\end{equation}
with $\widetilde{\Delta}$ and $\lambda$ warp factors to be defined below. The five-dimensional metric, which coincides with the metric in the lower-dimensional supergravity, is given by 
\begin{equation}
\dd s_5^2=\frac{L^2}{z^2\widetilde{X}^2}\left(\widetilde{X}^3\,\dd x_{1,3}^2+\dd z^2\right)\,.
\end{equation}
The constant $L$ is the radius of the $AdS$ in the UV of the flow, related to the coupling of the gauged supergravity by $g\,L=2^{2/3}$. The scalar $\widetilde{X}$ in Romans supergravity is related to the one in the $\mathcal{N}=4$ gravity multiplet as
\begin{equation}
\Sigma=\frac{1}{2^{1/3}}\frac{1}{\widetilde{X}}=\frac{1}{2^{1/3}}\frac{1}{\left(1+c_\sigma z^2\right)^{1/3}}\,.
\end{equation}
On the other hand, the internal metric is\footnote{The coordinate $\rho$ here is not the holographic radial coordinate, that we denoted $z$ above, but an internal one.}
\begin{equation}
\dd s_6^2=L^2\widetilde{X}\left\{\frac{\lambda^3}{4\left(1-\lambda^3\rho^2\right)}\left[\dd \rho^2+e^{D}\left(\dd y_1^2+\dd y_2^2\right)\right]+\frac{\lambda^3\rho^2}{4\widetilde{X}^2\widetilde{\Delta}}\dd\Omega_2^2+\frac{\left(1-\lambda^3\rho^2\right)\widetilde{X}}{\widetilde{\Delta}}\left(\dd y_3+V\right)^2\right\}\,.
\end{equation}
This metric is supported by the four-form which, in terms of the one-forms
\begin{align}
e^1=&\, \frac{L\,\lambda\,e^{D/2}}{2\left(1-\lambda^3\rho^2\right)^{1/2}} \,\dd y_1\,,  \qquad\qquad  e^2=\,    \frac{L\,\lambda\,e^{D/2}}{2\left(1-\lambda^3\rho^2\right)^{1/2}} \,\dd y_2\,,   \\[2mm]
e^3=&\,\frac{L\left(1-\lambda^3\rho^2\right)^{1/2}}{\lambda^{1/2}} \,\left(\dd y_3+V\right)\,, \qquad\qquad e^{4}=\, \frac{L\,\lambda}{2\left(1-\lambda^3\rho^2\right)^{1/2}}\,\dd\rho
\end{align}
and the two-form
\begin{equation}
\hat{\Omega}=\frac{L^2\,\lambda^2\,\rho^2}{4}\,\omega_2\,,
\end{equation}
with $\omega_2$ the volume form of the two-sphere, can be written as
\begin{equation}
F_4=-\frac{1}{\lambda^2\rho^2}\hat{\Omega}\wedge\dd\left[\widetilde{X}^{-2}\widetilde{\Delta}^{-1}\lambda^{1/2}\left(1-\lambda^3\rho^2\right)^{1/2}\rho \,e^3\right]-\frac{2}{L\lambda\rho}\hat{\Omega}\wedge e^{12}-\frac{2}{L\lambda^{5/2}\rho^2}\hat{\Omega}\wedge e^3\wedge e^4\,.
\end{equation}
The different warp factors in these expressions are given in terms of a solution to the Toda equation
\begin{equation}\label{eq:Toda}
\left(\partial_{y_1}^2+\partial_{y_2}^2\right)D+\partial_\rho^2e^D=0
\end{equation}
as
\begin{equation}
\lambda^3=-\frac{\partial_\rho D}{\rho\left(1-\rho\partial_\rho D\right)}\,,\qquad\qquad\widetilde{\Delta}=\widetilde{X}\lambda^3\rho^2+\widetilde{X}^{-2}\left(1-\lambda^3\rho^2\right)\,,
\end{equation}
while the fibration is
\begin{equation}
V=\frac12\partial_{y_2}D\,\dd y_1-\frac12\partial_{y_1}D\,\dd y_2\,.
\end{equation}
A particularly simple solution to \eqref{eq:Toda}, corresponding to \cite{Maldacena:2000mw}, is
\begin{equation}
e^{D}=\frac{1}{y_2^2}\left(\frac14-\rho^2\right)\,.
\end{equation}
Using this solution, the eleven-dimensional metric is regular everywhere except in the limit $z\to\infty$, which is the IR of the flow. In this limit, the five-dimensional metric goes as
\begin{equation}\label{eq:CB}
\dd s_5^2\sim\frac{L^2}{z^{4/3}}\left(-\dd t^2 + \dd\vec{x}^2+\frac{\dd z^2}{z^2}\right)\,,
\end{equation}
which cannot be put into hyperscaling form. In \cite{Freedman:1999gk} it was argued that this results in a continuous but gapped spectrum, and similar physics should apply to the alternative uplift \eqref{eq:CBuplift}. This is further elaborated in section \ref{sec:spectral}.


\section{Comments on the dual field theories}\label{sec:QFT}


In this section we aim to analyse some aspects of the QFTs dual to the various flows and fixed point solutions presented in sections \ref{sec:Nis2Theory}-\ref{sec:Nis4}. We only scratch the surface of a more in-depth study that will be the subject of future work.

As emphasised  throughout this paper, some of the solutions presented here are singular. Even when the singularities are ``good" or ``acceptable" according to 
standard criteria such as \cite{Gubser:2000nd,Maldacena:2000mw}, the singularities are indicating that some Physics is not being captured by the supergravity approximation\footnote{Even the comparatively well understood singularities associated to the presence of brane sources require the addition of degrees of freedom beyond those in supergravity.}. Nevertheless we expect the solutions presented here
to faithfully capture the dynamics of the QFT, if calculations are performed with the dual background sufficiently far from the singular locus. Such a perspective on singular supergravity backgrounds has proven to be valuable, as in the familiar cases of e.g. the Klebanov-Tseytlin \cite{Klebanov:2000nc}
and Klebanov-Strassler \cite{Klebanov:2000hb} backgrounds. Of course, we would ultimately like to find a mechanism by which one could resolve the singular behaviour. One might hope that this could be accomplished by allowing for a larger truncation of the ten or eleven-dimensional supergravity, an idea we have also left for future work.

We begin our analysis with a quick study of the linearised spectrum around the supersymmetric fixed point in eq.(\ref{bpsx}). We then briefly comment on the behavior of linearised 
fluctuations in the IR regime of eq.(\ref{solexcxx}), and the corresponding field theory expectations. 

Following this, we discuss an RG-monotone quantity commonly referred as ``central-" or ``$c$-function".
This quantity together with the Entanglement Entropy (for a rectangular region) and the rectangular Wilson loop serve as representative examples of a peculiar `decoupling' between the kinematical data defining the UV CFTs (ranks of colour and flavour groups, representations for matter fields) and the flow itself. In this sense, the flows described in sections  \ref{sec:Nis2Theory}-\ref{sec:Nis4} capture universal dynamics, applying to an infinite class of CFTs. 

\subsection{Linearised fluctuations around fixed points}

We can learn more about the Physics of the fixed points in eqs.(\ref{bpsx})-(\ref{nonbpsx}) by studying the spectra linearised around the solutions. This opens the possibility of flows connecting them. In particular, note that around the supersymmetric $AdS_5$, the linearised equations for the vectors can be written as
\begin{equation}
\dd\hf\Big( H^0+h H^1\Big) = 0 \qquad \mathrm{and} \qquad \dd\hf \Big(H^0-2h H^1\Big) = -\frac{6}{L^2}\hf\Big(A^0-2 h A^1\Big),
\end{equation}
which demonstrates that the vectors mix to form a massless and a massive vector mode, dual to operators of dimensions $\Delta = 3$ and $\Delta = 2 + \sqrt{7}$ respectively. The former is a conserved current, the latter an irrelevant vector operator in the dual CFT.

In the scalar sector, we have an $SO(2)$ doublet of operators dual to the $\theta^a$, with $m^2L^2= 3$. These are also irrelevant modes with $\Delta = 2+\sqrt{7}$. The scalars $\Sigma$ and $\varphi$ mix, and give rise to decoupled modes with $m^2L^2 = \mp 2(\sqrt{7}\mp2)$. These are relevant and irrelevant scalar operators, with dimensions $\Delta =1+\sqrt{7} $ and $\Delta = 3+ \sqrt{7}$ respectively.

As a consequence of supersymmetry, it follows that these dual operators organise into multiplets of $SU(2,2|1)$, the superconformal algebra of $\mathcal{N}=1$ SCFT's in four dimensions. Adapting the conventions of \cite{Cordova:2016emh}, one finds that in addition to the conserved stress tensor multiplet $A_1\bar{A}_1[1;1 ]^{r=0}_{\Delta = 3}$ (with $\Delta = 4$ stress tensor and $\Delta =3$ vector current) the remaining modes in the truncation fill out the long multiplet $L\bar{L}[ 0;0]^{r=0}_{\Delta = 1+\sqrt{7}}$. 

Although we will have little to say about the non-supersymmetric
critical point in this work, we note that the scalar modes retained in
our truncation are linearly stable around this vacuum. This extends
the analysis in \cite{Gauntlett:2002rv}, whose results we recover in the $\Sigma$, $\varphi$
sector.

\subsubsection{Spectral functions and their IR properties }\label{sec:spectral}

To understand the physical ramifications of the IR singularity represented in eq.(\ref{solexcxx}), it is helpful to consider the IR properties of spectral functions computed holographically in this background. In particular, we note that linearised fluctuations of the metric $h^x\,_y \sim e^{-i\omega t} h(\rho)$ obey a massless Klein-Gordon equation, whose most regular solution at the singularity is given by
\begin{equation}
h(\rho) = \sqrt{\rho}e^{\frac{\rho}{\sqrt{2}}\left(1-\sqrt{1-2\omega^2} \right)}\,U\left(\frac{3}{4}-\frac{1}{4\sqrt{1-2\omega^2}},\frac{3}{2},\sqrt{2}\rho\sqrt{1-2\omega^2}\right)
\end{equation}
where $U$ is the confluent hypergeometric function. This solution is manifestly real for frequencies $\omega <  1/\sqrt{2}$. It is thus expected on general grounds that the spectral function for the operator $T^x\,_y$ will exhibit a gap at low frequencies, with a continuous spectrum above \cite{Freedman:1999gk}.

\subsection{Holographic $c$-function}

In this section we shall study an interesting observable in Quantum Field Theory, the so-called ``$c$--function". At conformal fixed points, this observable evaluates to the central charge. It sometimes makes sense to associate a (varying) ``central charge" to flows between conformal points. In such cases, this quantity can provide a proxy for a  measure of the number of degrees of freedom (at a given energy scale) of the Quantum Field Theory, monotonically decreasing along the RG flow. 

Holographically, the central charge  of a $(d_s+1)$-dimensional conformal field theory is calculated as the volume of the `internal space'. That is the  $ (8-d_s)$-dimensional space complementary to the $AdS_{d_s+2}$ part of the geometry. This type of expression appeared first in the work of Brown and Henneaux (a precursor to the AdS/CFT correspondence) \cite{Brown:1986nw}. 

After that, the work of Freedman, Gubser, Pilch and Warner \cite{Freedman:1999gp} proposed a definition of the central charge that (as discussed above) was  basically the volume of the internal space. Their definition can be generalised to flows away from an $AdS$ fixed point. It was shown that such a quantity is indeed monotonic along the flow if certain energy conditions are imposed \cite{Freedman:1999gp}. Their results are particularly well suited to calculations directly in the lower-dimensional gauged supergravity. For a five-dimensional metric ansatz of the type studied in the present work, they define
\begin{equation}\label{eq:sugraCfn}
c(r) = \frac{\pi}{8G_5}\frac{1}{A'(r)^3}.
\end{equation}
It is simple to see that this expression holographically reproduces the central charge at a conformal fixed point, where $A' = L^{-1}$, with $L$ the $AdS$ radius. 

It is often advantageous to consider the higher dimensional origin of this formula. For generic field theories (like those describing the excitations of $Dp$-branes) the holographic definition of central charge was given in \cite{Klebanov:2007ws}. For the type of geometries we studied in this paper, the definition of \cite{Klebanov:2007ws} needs to be generalised as was done in \cite{Macpherson:2014eza}, \cite{Bea:2015fja}. In particular for field theories that flow across  dimensions the holographic definition was given in \cite{Bea:2015fja}.
\\
Let us review briefly the formalism developed in \cite{Klebanov:2007ws}, \cite{Macpherson:2014eza}, \cite{Bea:2015fja}. 

Consider a ten-dimensional background (in string frame) dual to a $(d_s+1)$ Quantum Field Theory of the form,
\begin{equation}
\dd s_{st}^2 = a(r,\vec{\theta})\left[\dd x_{1,d_s}^2 + b(r) \dd r^2 \right] + g_{ij}(r,\vec{\theta})\dd\theta^i \dd\theta^j.\label{metdiez}
\end{equation}
The background is complemented by a dilaton $\Phi(r,\vec{\theta})$. To calculate the central charge of the QFT we need to define the quantities
\begin{equation}
V_{int}=\int \dd\theta^i\sqrt{ e^{-4 \Phi}\det[g_{ij} ] a(r,\theta)^{d_s}}, \;\;\;\;\qquad H= V_{int}^2.\nonumber
\end{equation}
Then, following  \cite{Macpherson:2014eza}, \cite{Bea:2015fja}, we compute
\begin{equation}
c= \frac{d_s^{d_s}}{G_N} b(r)^\frac{d_s}{2} \frac{H^\frac{2d_s+1}{2}}{(H')^{d_s}}=\left(\frac{d_s}{2}\right)^{d_s}\frac{b(r)^{d_s/2}}{G_N} \frac{V_{int}^{(d_s+1)}}{(V_{int}')^{d_s}}.\label{centralch}
\end{equation}
Here, $'$ always denotes a derivative with respect to  the holographic coordinate, and $G_N$ is the ten dimensional Newton constant. If the background were given in Einstein frame, one can again use (\ref{centralch}), but this time with the quantities
\begin{equation}
V_{int}=\int \dd\theta^i\sqrt{\det[g_{ij} ] a(r,\theta)^{d_s}}, \;\;\;\;\qquad H= V_{int}^2.\nonumber
\end{equation}

If on the other hand we write the background as a solution of eleven-dimensional supergravity,
\begin{equation}
\dd s_{11}^2 =e^{-2\Phi/3} \left[a(r,\vec{\theta})\left(\dd x_{1,d_s}^2 + b(r) \dd r^2 \right) + g_{ij}(r,\vec{\theta})\dd\theta^i \dd\theta^j\right] + e^{4\Phi/3}(\dd x_{11}-A)^2,\label{meteleven}
\end{equation}
we calculate
\begin{equation}
V_{int}=\int \dd x_{11} \dd\theta^i \sqrt{ \det[g_{ij} ] a(r,\theta)^{d_s}}, \;\;\;\;\qquad H= V_{int}^2.\nonumber
\end{equation}
and we obtain exactly the same expression as in (\ref{centralch}), with the replacement $G_{N,10}=\frac{G_{N,11}}{\int \dd x_{11}}$.
\\
Let us now compute the $c$-function for the various geometries studied in this paper.

\subsection{$c$-Function along RG flows preserving four supercharges}

Consider  the metric for any of the supersymmetric backgrounds in section \ref{sec:Nis1flos}. We obtained these configurations as solutions to the BPS equations of a five-dimensional gauged supergravity. As explained in section \ref{sec:10n11dUp}, the eleven-dimensional metric reads
\begin{eqnarray}
& & \dd s_{11}^2= \Delta^{1/3} \dd s_{7}^2 +\frac{2}{\Delta^{2/3}} \dd s_4^2,\label{metricaonce}\\
& & \Delta= X^{-4} \sin^2\xi +\Delta \cos^2\xi,\;\;\;\;\qquad X^5= \Sigma^3 e^{-2\varphi},\nonumber\\
& & \dd s_{7}^2= (e^{\varphi} \Sigma)^{4/5}\left[e^{2A(r)} \dd x_{1,3}^2 + \dd r^2 \right] + (e^{\varphi}\Sigma)^{-6/5} \dd\Omega_{2,l},\nonumber\\
& &  \dd s_{4}^2= X^3\Delta \dd\xi^2+ \frac{\cos^2\xi}{4 X}\left[\dd\alpha^2+\sin^2\alpha \dd\beta^2+ (\dd\psi -A^{3})^2 \right].\nonumber
\end{eqnarray}
Explicitly, we can choose coordinates on the Riemann surface such that the vector $A^{3}= \cosh y \,\dd v$ (the spin connection of $H^2$) for $\dd\Omega_{2,l=-1}= \dd y^2+\sinh^2 y\, \dd v^2$. Similarly, for $l=1$
 we have $A^{3}= \cos\theta\, \dd\phi$ (the spin connection on $S^2$) and $A^{3}=0$ for $l=0$.
 \\
 Following the nomenclature of eqs.(\ref{metdiez})-(\ref{meteleven}) and considering that we aim to calculate the $c$-function of the dual four dimensional QFT, we find
 \begin{eqnarray}
& & d_s=3, \;\;\;\;\qquad a= \Delta^{1/3} (e^{\varphi}\Sigma)^{4/5} e^{2 A(r)},\;\;\;\;\qquad b= e^{-2A(r)},\nonumber\\
& & \dd s_{int}^2= \Delta^{1/3}  (e^{\varphi}\Sigma)^{-6/5} \dd\Omega_{2,l} +\frac{2}{\Delta^{2/3}} \left( X^3\Delta \dd\xi^2+ \frac{\cos^2\xi}{4 X}\left[\dd\alpha^2+\sin^2\alpha \dd\beta^2+ (\dd\psi -A^{3})^2 \right] \right),\nonumber\\
& & \det{g_{int}}\, a^3= e^{6A} \frac{\cos^6\xi \sin^2\alpha }{4}\mathrm{vol}_{2,l}^2.\nonumber
\end{eqnarray}
 Evaluating (\ref{centralch}), one obtains
 \begin{eqnarray}
 & & c= \frac{{\mathbb{N}}}{8 G_{N,11}} \frac{1}{A'(r)^3},\\
 & & 4 {\mathbb N}= \int_0^{\pi/2} \cos^3\xi  \dd\xi \int_{0}^\pi \sin\alpha\, \dd\alpha \int_{0}^{2\pi} \dd\beta \int_{0}^{4\pi} \dd\psi \int_{\Omega_{2,l}} \dd\mathrm{vol}_{2,l}\nonumber
 \end{eqnarray}
 In perfect agreement with (\ref{eq:sugraCfn}).
 
 A more explicit expression for $A'(r)$ can be found by using the BPS equations (\ref{eqAmain})-(\ref{eqvarphimain})
and the material  and definitions of Appendix \ref{eqs5dappendix}. We find in terms of the `coordinate' $\Sigma$ and the function $y(\Sigma)$,
\begin{equation} 
 A'(r)= \frac{1}{y(\Sigma)}\left( \frac{f_4 + e^{2\varphi} f_5}{f_1+ e^{2\varphi} f_2}\right),\;\;\;\;\qquad e^{2\varphi}= \frac{1- y(\Sigma) f_1}{y(\Sigma) f_2}.
 \end{equation}
%
%

\subsubsection{The flow and its uplift to Massive IIA}

 Some field-theoretical aspects of the flow become clearer when analysing the lift of the five-dimensional supergravity solution to Massive IIA.
Let us briefly recall the lift to Massive IIA  described in section \ref{liftmassiveiia}. For the flow we are interested in, and referring to eq.(\ref{lift10xxx}), we have that 
 \begin{eqnarray}
 & & \dd s_{7}^2= (e^{\varphi} \Sigma)^{4/5}\left[e^{2A(r)} \dd x_{1,3}^2 + \dd r^2 \right] + (e^{\varphi}\Sigma)^{-6/5} \dd\Omega_{2,l},\;\;\;\qquad X^5= \Sigma^3 e^{-2\varphi},\nonumber\\
 & & D\Omega_2= \dd\chi^2+\sin^2\chi (\dd\psi- \frac{1}{2} A^{3})^2.\nonumber
\end{eqnarray}
%

 We can compute a holographic $c$-function following a procedure similar to the one outlined above. We have, for a solution describing an RG
flow of a four-dimensional field theory,
  \begin{eqnarray}
& & d_s=3, \;\;\;\; a=  X^{-1/2} (e^{\varphi}\Sigma)^{4/5} e^{2 A(r)} \sqrt{-\frac{\alpha}{\ddot{\alpha}}} ,\;\;\;\; b= e^{-2A(r)},\nonumber\\
& & \dd s_{int}^2= X^{-1/2} \sqrt{-\frac{\alpha}{\ddot{\alpha}}} (e^{\varphi}\Sigma)^{-6/5} \dd\Omega_{2,l} +  X^{5/2}\sqrt{-\frac{\ddot{\alpha}}{\alpha}}\left[\dd z^2 +\frac{\alpha^2}{\dot{\alpha}^2 - 2 \alpha \ddot{\alpha} + 2 \alpha \ddot{\alpha} (1-X^5)} D\Omega_2\right].\nonumber\\
& & e^{-4 \Phi}\det{g_{int}} a^3=  e^{6A} (-\alpha \ddot{\alpha})^2 \sin^2\chi\, \mathrm{vol}_ {2,l}^2.\nonumber
\end{eqnarray}
 Using this data in (\ref{centralch}) yields
 \begin{eqnarray}
 & & c= \frac{{\mathbb N}}{8 G_{N,10}} \frac{1}{A'(r)^3},\\
 & &  {\mathbb N}= \int_{0}^\pi \sin\chi \,\dd\chi \int_{0}^{2\pi} \dd\psi \int_{\Omega_{2,l}} \dd\mathrm{vol}_{2,l} \int_{0}^{z_*} (-\alpha \ddot{\alpha}).\label{centralflow}
 \end{eqnarray}
Notice that the coefficient ${\mathbb N}$ is proportional to that obtained in \cite{Nunez:2018ags} when calculating the central charge of the six-dimensional CFTs. The interpretation of (\ref{centralflow}) is then clear. Flowing away from the CFT, the $c$-function factorises into a component coming purely from the original conformal point (encoded in ${\mathbb N}$, and dependent only on the original CFT and on the volume of the compactification manifold)
 and  a contribution from the flow itself, captured by $A'(r)$. In other words, the dynamics of the RG flow is independent of the data defining the SCFT. The same RG flow is universal for all the SCFTs with a trustable string dual.
 
 \subsection{$c$-function along flows preserving eight supercharges}
 
 Consider now the flow discussed in section \ref{sec:8Qflo}. More particularly, we focus on the interesting special case of the background described below (\ref{eq:CBuplift}). In the coordinates there, we have an eleven-dimensional metric of the form\footnote{To avoid cluttering the formulas, here we shall denote by $X,\Delta$ what below eq.(\ref{eq:CBuplift}) was denoted by $\widetilde{X},\widetilde{\Delta}$.},
 
 \begin{eqnarray}
 & & \dd s_{11}^2= \frac{1}{4}\left( \frac{\Delta X^2 L^6}{\lambda^3}\right)^{1/3} \left[ \frac{4}{z^2}(\dd x_{1,3}^2+\frac{\dd z^2}{X^3})  + \dd s_{int}^2 \right],\label{metriconcegm}\\
  & & \dd s_{int}^2=\frac{\lambda^3}{1-\lambda^3 \rho^2}\left(\dd\rho^2+ e^{D}(\dd y_1^2+ \dd y_2^2)  \right) + \frac{\lambda^3 \rho^2}{\Delta X^2} \dd\Omega_2 + (1-\lambda^3 \rho^2)\frac{X}{\Delta}(\dd y_3-V)^2\nonumber\\
  & & \lambda^3= -\frac{\partial_\rho D}{\rho(1- \rho \partial_\rho D)}, \;\;\;\;\; D=D(\rho, y_1,y_2),\nonumber\\
  & &\Delta= \lambda^3 \rho^2 \frac{(X^3-1)}{X^2} +\frac{1}{X^2},\;\;\;\; X=X(z).\nonumber
  \end{eqnarray}
When the field $X(z)=1$ we have $\Delta=1$. The function $\lambda (y_1,y_2,\rho)$ is defined in terms of the function $D(y_1,y_2,\rho)$ which satisfies a Toda equation. In the particular case in which the solution to the Toda equation has an extra isometry in the $y_1$-$y_2$ plane, we can reduce to Type IIA. The interpretation in terms of a long linear quiver was given in \cite{Gaiotto:2009gz}-\cite{Nunez:2019gbg}. 

A straightforward calculation of the $c$-function in this case leads to
\begin{eqnarray}
& & \det[g_{int}] a^3= \frac{L^{18}}{4^5} \frac{e^{2D} \lambda^6 \rho^4 X^3 \sin^2 \chi}{(1-\lambda^3\rho^2)^2 z^6},\nonumber\\
& & V_{int}=\int \dd\rho \dd y_1 \dd y_2 \dd y_3 \dd\chi \dd\xi \sqrt{\det g_{int} a^3 }= {\mathbb N}\frac{X^{3/2}}{z^3}.
\end{eqnarray}
Under the (non-essential) additional assumption that $D=D(\rho)$, we can further evaluate
\begin{equation}
{\mathbb N}= \frac{L^9}{32} V_{y_1y_2} \times 2\pi \times 4\pi \int_{0}^{\rho_*}\dd\rho \frac{e^{D} \lambda^3 \rho^2}{1-\lambda^3 \rho^2}
=- \frac{\pi^2 L^9}{4}V_{y_1y_2} \int_{0}^{\rho_*} \dd\rho\, \rho \partial_\rho e^{D}.
\end{equation}

 Following the procedure outlined above we have
 \begin{equation}
 H= {\mathbb N}^2 \frac{X^3}{z^6}, \;\;b(z)= \frac{1}{X^3}, \;\;\;\; c= \frac{{\mathbb N}}{8 G_{N,11}} \frac{1}{(-X(z) + \frac{z X'(z)}{2})^3} = -\frac{{\mathbb N}}{8 G_{N,11}L^3} \frac{1}{A'(r)^3}.
 \end{equation}
  In the final equality, we have used the BPS equations and (\ref{eq:drdzNis4}) with $\varphi_i = 0$ to bring the expression to a more familiar form. 
  
 We find here the same phenomenon as  above. Namely, when evaluated along these flows, the $c$-function factorises into a part determined solely by the quiver structure (in particular the colour and flavour groups of the original CFT) encoded by ${\mathbb N}$, and a part controlled by the details of the flow as the radial coordinate changes. In other words, the same kind of universality as found above is at work here. The RG-flow holographically described by the BPS equations (\ref{eqAmain})-(\ref{eqvarphimain}) is the same for all the Gaiotto SCFTs with a trustworthy string dual.
 
\subsection{Entanglement Entropy and Wilson loops}

Finally, we briefly comment on the higher-dimensional perspective of the holographic computation of strip entanglement entropy and Wilson loops in the field theory phases dual to the backgrounds we presented in this paper. For brevity, we shall study only  the five-dimensional backgrounds in section \ref{sec:Nis1flos}  when lifted to Massive IIA as indicated in section \ref{liftmassiveiia}. The notable feature stressed above, namely the ``factorisation"  between the CFT data and the dynamics imposed by the flow, will also appear here (as well as in the other backgrounds in eleven dimensions). The result in this sense is again universal.

The holographic prescription for computing an entanglement entropy was first given in \cite{Ryu:2006ef}. For our backgrounds, which describe flows away from a fixed point, we find it convenient to use the methods developed by \cite{Klebanov:2007ws}. To calculate the entanglement entropy between two rectangular regions separated a distance $\bar{L}$, we consider an eight-manifold defined by $M_8=[x_1,x_2,x_3, H_2, z, \Omega_2]$. There is a dependence of the holographic coordinate $r=r(x_1)$. The induced (string frame) metric on such manifold is
\begin{eqnarray}
& & \frac{1}{\mu}\dd s_{8}^2=X^{-1/2}\sqrt{-\frac{\alpha}{\ddot{\alpha}}}\left[ (\Sigma e^{\varphi})^{4/5} e^{2A}\left(\dd x_2^2+\dd x_3^2 + \dd x_1^2(1+ e^{-2A} r'^2)  \right)  + (\Sigma e^{\varphi})^{-6/5} \dd\Omega_{2,l} \right]  \nonumber\\
& &+ X^{5/2}\sqrt{-\frac{\ddot{\alpha}}{\alpha}}(\dd z^2 + \frac{\alpha^2}{N} D\Omega_{2}).\nonumber
\end{eqnarray}
Here $N=\dot{\alpha}^2 - 2 \alpha \ddot{\alpha} + 2 \alpha \ddot{\alpha} (1-X^5)$.
Following \cite{Klebanov:2007ws} we calculate
\begin{eqnarray}
& & S_{EE}=\frac{1}{G_{N,10}}\int_{M_8}e^{-2\Phi} \sqrt{\det[g_{ind}]}={\mathbb M} \int_{0}^{\bar{L}}\dd x_1 e^{3 A(r)}\sqrt{1+ e^{-2A(r) }  r'(x_1)^2},\label{eedef}
\end{eqnarray}
with 
\begin{equation}
G_{N,10 }\, {\mathbb M}=e^{-2\Phi_0} \mathrm{vol}_{x_1,x_2} \mathrm{vol}_{2,l} \,\mathrm{vol}_{S^2} \int_0^{z_*} (-\alpha \ddot{\alpha}).\nonumber
\end{equation}
Again, we find a decoupling between the dynamics 
of the observable induced by the flow--represented by the integral in (\ref{eedef}), and the UV CFT data encoded in ${\mathbb M}$. The computation proceeds by dealing with a minimisation  problem as described in \cite{Kol:2014nqa}.

Along very similar lines, we can calculate the Wilson loop defined in \cite{Maldacena:1998im}.  To holographically  compute the energy between a pair of heavy quarks in the field theory that are separated a distance $\bar{L}$ in time $T$, we consider a string in the configuration
$t=\tau,x_1=\sigma$ with $r=r(\sigma)$ and $z=z_0$ fixed. We calculate the induced metric and the Nambu--Goto Action for this string
\begin{eqnarray}
& & \dd s_{ind}^2= X^{-1/2} \sqrt{ -\frac{\alpha}{\ddot{\alpha} } } (\Sigma e^{\varphi})^{4/5} e^{2A}\left[ -\dd t^2 + \dd\sigma^2(1+ e^{-2 A} r'^2)\right],\label{wilsonlxx}\\
& & S_{NG}= \frac{1}{2\pi\alpha'} \int \dd\tau \dd\sigma\sqrt{-\det{g_{ind} } }=  {\mathbb T}\int_0^{\bar{L}} \dd\sigma F(r(\sigma)) \sqrt{(1+ e^{-2 A} r'^2)},\nonumber\\
& & 2\pi\alpha' {\mathbb T}= T \sqrt{\frac{\alpha(z_0)}{\ddot{\alpha} (z_0)}},\;\;\;\; F^5= \Sigma e^{6\varphi}.\nonumber
\end{eqnarray}
Unlike the case of the $c$-function and the strip entanglement entropy considered above, the dynamics of this particular string probe is independent of the detailed CFT data---see \cite{Kol:2014nqa} for a qualitative analysis of the different behaviour of Entanglement Entropy and Wilson loop. 

The calculation proceeds along the usual line, by minimising the Nambu--Goto action. Although we do not perform this calculation here, we can nonetheless comment on our expectations for this observable in the states dual to our solutions. As noted above in discussing solutions to the massless Klein-Gordon equation in the IR geometries, these flows share important qualitative features with the ``Coulomb branch" flow found in \cite{Pilch:2000ue}. Holographic spectral functions and Wilson loops were studied in this background in \cite{Freedman:1999gk}. There it was argued that the nature of the IR singularity is such that the dual phase is gapped, but not confining. In particular, the Wilson loop exhibits ``perfect screening" at large distances, characterised by perimeter law behaviour. We expect a similar behaviour for our QFTs. This is another universal aspect of the holographic QFTs defined in this work.


\section{Discussion, conclusions and future work}\label{concl}


Let us start with a brief summary.
In this work we have exploited the power of consistent Kaluza--Klein truncations of supergravity to study 1/2-BPS RG flows in a large class of four-dimensional superconformal field theories in the holographic limit. Included in this class are certain examples of the $\mathcal{N}=2$ class-$\mathcal{S}$ theories of \cite{Gaiotto:2009we}, as well as their massive deformations described by the  $\mathcal{N}=1$ ``Sicilian" gauge theories introduced in \cite{Benini:2009mz}. 

As a consequence of the plurality of the available uplifts to higher-dimensional supergravity (either to eleven dimensions or Massive IIA), these same five-dimensional solutions can in fact describe supersymmetry preserving RG flows in infinite families of four-dimensional SCFTs. This provides yet another interesting realisation of universality in strongly coupled holographic field theories. Along this line, qualitatively similar universalities were observed in \cite{Bobev:2017uzs}.
We have studied certain aspects of the dual field theories, like those encoded in linearised fluctuations around particular backgrounds.  Importantly we described `universal' quantities, for which the data defining the UV CFT 
decouples from the details of the flow.

Our results encourage a number of interesting directions in which future efforts could be focused. One example is the landscape of {\it non}--supersymmetric RG flows in these SCFTs. As indicated in section \ref{sec:Nis2Theory}, when $l = -1$, the five-dimensional theory has two $AdS_5$ solutions. Only one of these is supersymmetric, but the two are in fact quite close to one another in field space. For example, their radii satisfy
\begin{equation}
\frac{L_{\mathrm{susy}}}{L_{\mathrm{nosusy}}} \approx 1.00009.\nonumber
\end{equation}
It seems likely that a flow exists from the supersymmetric fixed point to the fixed point in which all supersymmetry is broken. To construct this flow, it will be necessary to solve the (second order) equations of motion, which can be accomplished numerically using standard techniques. 

Alternatively, it may be interesting to look more closely at flows for other values of $l$. These flows do not terminate at an $AdS_5$, but could nonetheless be phenomenologically novel and physically relevant. For example, when $h=0$ and $l=1$, the BPS equations are solved by the well-known Chamseddine-Volkov flow \cite{Chamseddine:1997nm}, which is of interest because its IR behaviour is holographically interpreted as the confining phase of strongly-interacting ${\cal N}=1$ Super Yang-Mills coupled to adjoint matter \cite{Maldacena:2000yy}.  It would be interesting to learn whether or not a flow exists in the $h\ne 0$ theory which is described in the UV by an $AdS_7$ solution with two spatial directions curled into an $S^2$, and in the IR by the Chamseddine--Volkov solution.

Moreover, black hole solutions corresponding to finite temperature states of these theories could be constructed, cloaking the singularities and thus showing explicitly that they are good in the sense of \cite{Gubser:2000nd}. Some peculiarities are to be expected in the thermodynamic properties of such black holes in analogy to those of the Coulomb branch solution \cite{Myers:2001aq}.

Another avenue that may be worth pursuing is a more thorough investigation of the brane configurations which give rise to the RG flow solutions we have found. In many respects, the dual physics of these solutions is reminiscent of that of a Coulomb branch solution in $\mathcal{N}=4$ Super Yang--Mills \cite{Pilch:2000ue} (or the analogous flow in the ABJM theory \cite{Pope:2003jp,Bena:2004jw}). In both these examples, the higher dimensional geometries reveal that the supergravity solutions are supported by ``continuous distributions" of branes \cite{Freedman:1999gk}. It would be interesting to see what can be learned in the present case. 

It would also be interesting to repeat the analysis in this work, but for lower-dimensional gauged supergravities, with a view towards constructing flows away from SCFTs described by the backgrounds in \cite{DHoker:2016ujz}-\cite{Couzens:2019mkh}. 

Finally, we highlight two future directions in which we have already taken some preliminary steps. The first is an investigation of 1/4-BPS flows in the $\mathcal{N}=4$ theory, using the results of \cite{Cassani:2012wc}. The second is to expand the $\mathcal{N}=2$ truncation of section \ref{sec:Nis2Theory} by embedding it in the maximal theory in seven dimensions. Broadly, from the discussion in \cite{Benini:2009mz} one might anticipate an enlargement to an $\mathcal{N}=2$ theory whose gauge group can be extended to include $U(1)_R\times SU(2)_F$ as a compact subgroup. From the seven dimensional perspective, this should be the $U(1)_R\times SU(2)_F$ inside the $SO(4)\subset SO(5)$ gauge group (also anticipated in \cite{Cassani:2019vcl}). We hope to report on these developments shortly.

\subsection*{Acknowledgements}

We thank  Davide Cassani, Jerome Gauntlett for helpful discussions, as well as Matthew Cheung for collaboration on closely related themes. AF is supported by grants FPA2016-76005-C2-1-P, FPA2016-76005-C2-2-P, SGR-2017-754 and MDM-2014-0369. The work of CR is funded by a Beatriu de Pin\'os Fellowship.

\begin{appendix}

\section{Details of the reduction}\label{sec:eomsA}
\subsection{Equations of motion}

The equations of motion of minimal gauged supergravity in $D=7$ are
\begin{align}
0 & = R_{MN} - 5 X^{-2}\nabla_M X \nabla_N X-\frac{1}{5}\mathcal{V}g_{MN}-\frac{1}{2 X^2}[F_{(2)}]^2_{MN}-\frac{X^4}{2}[F_{(4)}]^2_{MN}\\
0 & = \nabla_M(X^{-1}\nabla^M X)-\frac{X^4}{120}F_{(4)}^2+\frac{1}{20 X^2} F_{(2)}^2-\frac{X}{10}\partial_X\mathcal{V}\\
0 & = \dd (X^4\hs F_{(4)})+2h F_{(4)}-\frac{1}{2} F_{(2)}^i\wj F_{(2)}^i\\
0 & = D (X^{-2}\hs F^i_{(2)})-F^i_{(2)}\wj F_{(4)}
\end{align}
where we introduced the $SU(2)$ covariant derivative $D$ which acts on an $SU(2)$ triplet as e.g. $D H^i = \dd H^i -\as^{ijk} A_{(1)}^j\wj H^k$, and further defined
\begin{equation}
[F_{(2)}]^2_{MN} \equiv F_{(2)}^i\,_{MP}F_{(2)}^i\,_N\,^P-\frac{1}{10}F_{(2)}^2 g_{MN}, \qquad [F_{(4)}]^2_{MN} \equiv \frac{1}{6}F_{(4)}\,_{MPQR}F_{(4)}\,_N\,^{PQR}-\frac{1}{40} F_{(4)}^2 g_{MN}.
\end{equation}

For non-vanishing topological mass $h$, it is necessary to impose the first order constraint to perform a correct accounting of the degrees of freedom. To motivate the form of the ansatz for the three-form, it is thus helpful to note that in this reduction
\begin{multline}
\frac{1}{2} A^i_{(1)}\wj F^i_{(2)}+\frac{1}{12}\as_{ijk} A^i_{(1)}\wj A^j_{(1)}\wj A^k_{(1)} = \\
-\frac{1}{2}\dd\left(\bar{\omega}^{12}\wj\mathcal{A} \right)+\frac{1}{2}\mathcal{A}\wj\mathcal{F}+l\mathcal{A}\wj\mathrm{vol}_2-\as^{ab}\theta^a D\theta^b\wj \mathrm{vol}_2.
\end{multline}
From the ansatz for the three-form potential
\begin{equation}
B_{(3)} = c_{(3)}+\chi_{(1)}\wj \mathrm{vol}_2 -\frac{1}{2h}\dd(\bar{\omega}^{12}\wj\mathcal{A}),
\end{equation}
it follows that
\begin{equation}
F_{(4)} = f_{(4)}+\dd\chi_{(1)}\wj\mathrm{vol}_2 \qquad \mathrm{and}\qquad \star F_{(4)} = e^{12\phi}\hf f_{(4)}\wj\mathrm{vol}_2 + e^{-8\phi}\hf \dd\chi_{(1)},
\end{equation}
where $f_{(4)} = \dd c_{(3)}$. The self-duality constraint yields the relations
\begin{align}
X^4 e^{12\phi} \hf f_{(4)} & = -2h\, \chi_{(1)}+l\mathcal{A}-\as^{ab}\theta^a D\theta^b, \label{eq:rcon1}\\
X^4 e^{-8\phi}\hf \dd\chi_{(1)} & = -2h\, c_{(3)} + \frac{1}{2}\mathcal{A}\wj \mathcal{F}.\label{eq:rcon2}
\end{align}
 
 The reduction of the flux equations of motion yields an equation of motion for the doublet of scalars $\theta$,
\begin{equation}
0 = D\left( X^{-2}e^{-6\phi}\hf D\theta^a\right)+X^{-2}e^{-16\phi}\theta^a(l-\theta\cdot\theta)\hf 1-\as^{ab}D\theta^b\wj f_{(4)},
\end{equation}
a Maxwell equation for the vector $\mathcal{A}$,
\begin{equation}
0 = \dd\left(X^{-2}e^{4\phi}\hf \mathcal{F} \right)+2 X^{-2}e^{-6\phi}\as^{ab}\theta^a\hf D\theta^b-\mathcal{F}\wj\dd\chi_{(1)}-(l-\theta\cdot\theta)f_{(4)}
\end{equation}
and second order equations for the modes appearing in $F_{(4)}$,
\begin{align}
0 & = \, \dd \Big(X^{4}e^{12\phi}\hf f_{(4)}+2h\chi_{(1)}-l\mathcal{A}+\as^{ab}\theta^a D\theta^b \Big),\\
0 & = \, \dd \Big(X^{4}e^{-8\phi}\hf \dd \chi_{(1)}+2h c_{(3)}-\frac{1}{2}\mathcal{A}\wj \mathcal{F} \Big).
\end{align}
These have been written so as to emphasise the fact that the reduced second order equations are compatible with the reduced constraints (\ref{eq:rcon1}) and (\ref{eq:rcon2}), as anticipated.

Introducing for convenience the two-form field strength $\mathcal{H} = \dd \chi_{(1)}$, the reduction of the Klein-Gordon equation for $X$ gives the five-dimensional equation
\begin{multline}
0 = \dd \left(X^{-1}\hf \dd X \right)-\frac{1}{5}X^4\left(e^{12\phi}f_{(4)}\wj\hf f_{(4)}+e^{-8\phi}\mathcal{H}\wj\hf\mathcal{H} \right)\\+\frac{1}{10}X^{-2}\left(2e^{-6\phi}D\theta^a\wj\hf D\theta^a + e^{4\phi}\mathcal{F}\wj\hf\mathcal{F}+e^{-16\phi}(l-\theta\cdot\theta)^2\hf 1 \right)\\
+\frac{2}{5}X^{-8}\left(4h^2-3\sqrt{2}h X^{5}+X^{10} \right)\hf 1,
\end{multline}

while the reduced Einstein equations give a Klein-Gordon equation for $\phi$ 
\begin{multline}
0 = \nabla^2\phi -\frac{1}{3}l e^{-10\phi}+\frac{1}{15}e^{-4\phi}\mathcal{V}-\frac{1}{2X^2}\left[\frac{1}{30}e^{4\phi}\mathcal{F}^2-\frac{1}{5}e^{-6\phi}(D\theta^a)^2-\frac{4}{15}e^{-16\phi}(l-\theta\cdot\theta)^2 \right]\\
-\frac{1}{2}X^4\left[ \frac{1}{120}e^{12\phi} f_{(4)}^2 -\frac{1}{15}e^{-8\phi}\mathcal{H}^2 \right]
\end{multline}
and five-dimensional Einstein equations
\begin{multline}
0 = \bar{R}_{mn}-30\nabla_m\phi\nabla_n\phi-5 X^{-2}\nabla_m X\nabla_n X-\frac{1}{X^2}e^{-6\phi}D_m\theta^a D_n\theta^a\\
-\frac{1}{3}\left(e^{-4\phi}\mathcal{V}-2le^{-10\phi}+\frac{1}{2X^2}e^{-16\phi}(l-\theta\cdot\theta)^2 \right)\bar{\eta}_{mn}\\
-\frac{1}{2 X^2}e^{4\phi}\mathcal{F}_{mp}\mathcal{F}_n\,^p+\frac{1}{12 X^2}e^{4\phi}\mathcal{F}^2\bar{\eta}_{mn}-\frac{1}{2}X^4 e^{-8\phi}\mathcal{H}_{mp}\mathcal{H}_n\,^p+\frac{1}{12}X^4e^{-8\phi}\mathcal{H}^2\bar{\eta}_{mn}\\
-\frac{1}{12}X^4 e^{12\phi}f_{mpqr}f_n\,^{pqr}+\frac{1}{48}X^4e^{12\phi}f_{(4)}^2\bar{\eta}_{mn}.
\end{multline}

It is straightforward to show that these reduced equations of motion, taken together with (\ref{eq:rcon1}, \ref{eq:rcon2}), are equivalent to the equations of motion of the $\mathcal{N}=2$ theory described by (\ref{eq:N2axn}).

\subsection{Special geometry}\label{sec:geoA}
In our reduced theory, the manifold $SO(1,1)$ is characterised by the symmetric tensors
\begin{equation}
\mathcal{C}_{100}=\frac{\sqrt{3}}{2}, \qquad a_{00} = \Sigma^{-2}, \quad a_{11} = \Sigma^4
\end{equation}
and can be embedded in a two-dimensional space spanned by
\begin{equation}
h^{0} =\sqrt{\frac{2}{3}}\Sigma \qquad \mathrm{and} \qquad h^1 = \frac{1}{\sqrt{3}}\Sigma^{-2} \qquad \mathrm{with}\qquad \mathcal{C}_{\tilde{I}\tilde{J}\tilde{K}}h^{\tilde{I}}h^{\tilde{J}}h^{\tilde{K}} = 1.
\end{equation}
Consistency requires
\begin{equation}
a_{\tilde{I}\tilde{J}} = -2\mathcal{C}_{\tilde{I}\tilde{J}\tilde{K}}h^{\tilde{k}}+3h_{\tilde{I}}h_{\tilde{J}} \qquad \mathrm{and} \qquad g_{xy} = 3\partial_x h^{\tilde{I}}\partial_y h^{\tilde{J}} a_{\tilde{I}\tilde{J}},
\end{equation}
which is indeed the case.

To understand our paramerisation of $SU(2,1)/U(2)$, we mostly follow the conventions of \cite{Cheung:2019pge}. The metric can be written in an orthonormal frame such that $g_{UV} = f^a\,_U f^a\,_V$, using the vierbeins
\begin{equation}
f^1 = 2 \dd\varphi, \quad f^2 = e^{2\varphi}\Big(\dd \xi-\as^{ab}\theta^a\dd \theta^b \Big), \quad f^3 = \sqrt{2}e^{\varphi}\dd\theta^1, \qquad f^4 = \sqrt{2}e^{\varphi}\dd\theta^2.
\end{equation}
The $SO(4)$ valued spin connection is given by
\begin{equation}
\omega = \frac{1}{2}\Big[\Big(2 M_{21}+M_{34} \Big)f^2 +\Big(M_{31}+M_{24} \Big)f^3 + \Big( M_{41}+M_{32}\Big) f^4\Big],
\end{equation}
where $M_{mn} = E_{mn}-E_{nm}$ generate $SO(4)$--the matrix $E_{mn}$ has a 1 in the $m, n$ positions, and zeroes elsewhere. Decomposing the spin connection into $SU(2)\times Sp(2)$ it is straightforward to extract the complex structures valued in $SU(2)$. They are
\begin{equation}
\vec{J} = -\Big(f^{14}+f^{23},f^{24}-f^{13},f^{12}+f^{34} \Big).
\end{equation}
The corresponding moment maps are then 
\begin{align}
\vec{P}_0 & = \, \Big(\sqrt{2}e^{\varphi}\theta^1,\sqrt{2}e^{\varphi}\theta^2,1+\frac{1}{2}e^{2\varphi}(l-\theta\cdot\theta) \Big),\\
\vec{P}_1 & = \, \Big(0,0,h e^{2\varphi} \Big),
\end{align}
and with (\ref{eq:Vs}-\ref{eq:vPI}) one arrives at a scalar potential given by
\begin{equation}
V_s = -2\Sigma^2-4\sqrt{2}e^{2\varphi}\Sigma^{-1} h + 2e^{4\varphi}\Sigma^{-4} h^2-2 l e^{2\varphi}\Sigma^2+\frac{1}{2}e^{4\varphi}\Sigma^2 (l-\theta\cdot\theta)^2,
\end{equation}
matching exactly that of the reduced theory.

\subsection{Comment on the $h=0$ limit}\label{sec:his0}
The reduction applies equally well to the case of vanishing topological mass in seven dimensions. Operationally, the only non-trivial step is to replace the reduction ansatz for the three form with
\begin{equation}
B_{(3)} = c_{(3)}+\chi_{(1)}\wj \mathrm{vol}_2.
\end{equation}
In other words, the singular-in-the-limit exact term is no longer present, as the three form now satisfies a conventional second order equation of motion (and such a term is pure gauge). The broadstroke features of the analysis continue as before, and one eventually obtains (\ref{eq:N2axn}) with $h=0$.

An interesting consequence of the massless limit is that the five-dimensional theory acquires a more intricate gauging. This gauging is now entirely along the Killing vector
\begin{equation}
k_0 = \theta^1\partial_{\theta^2}-\theta^2\partial_{\theta^1}+l\partial_\xi
\end{equation}
which generates an isometry inside of $SU(2,1)/U(2)$. For the case of toroidal reduction with $l=0$ one finds that the gauge group is compact, $SO(2)$, whereas for reductions on the sphere or hyperbolic plane the resulting gauge group is a non-compact mixture with an $\mathbb{R}'\subset SO(2)\times \mathbb{R} \subset SU(2,1)/U(2)$.

When $h=0$, there is the possibility to include an additional scalar mode in the BPS equations. The scalar potential can be derived from the superpotential
\begin{equation}
W_{h=0} = \frac{\Sigma}{3\sqrt{2}}\sqrt{4+e^{4\varphi}\big(l-\theta\cdot\theta\big)^2+4e^{2\varphi}\big(l+\theta\cdot\theta\big)}
\end{equation}
and the flow equations are given by
\begin{align}
\partial_r A & = W,\\
\partial_r\Sigma & = -\Sigma W,\\
\partial_r\varphi & = -\frac{1}{6W}\Sigma^2\Big(e^{2\varphi}\big(l-\theta\cdot\theta\big)^2+2\big(l+\theta\cdot\theta\big) \Big)e^{2\varphi},\\
\partial_r\theta^a & = -\frac{1}{3W}\Sigma^2\Big(2-e^{2\varphi}\big(l-\theta\cdot\theta) \Big)\theta^a.
\end{align}

Solutions to this $h=0$ gauged supergravity in $d=5$ can thus be uplifted through $D=7$ to both Type I supergravity in $D=10$ and to the $D=11$ theory via the results of \cite{Chamseddine:1999uy}. This uplift uses the formalism in which the $D=7$ three-form has been dualised, which is only available in the $h=0$ case. As emphasised in their work,  the eleven-dimensional solutions are thus related to the lift of the $h\ne0$ solutions through a singular limit.


\section{Working with the 5d BPS equations}\label{eqs5dappendix}


In this appendix we  discuss analytic manipulations on the system of  BPS equation found  in eqs.(\ref{eqAmain})-(\ref{eqvarphimain}). We begin by writing the equations,
\begin{eqnarray}
& & \Sigma'=f_1(\Sigma) +  {e^{2\varphi}} f_2(\Sigma),\label{eqS}\\
& & \varphi'=  \frac{e^{2\varphi}}{2} f_3(\Sigma),\label{eqvarphi}\\
& & A'= f_4(\Sigma) + e^{2\varphi} f_5(\Sigma)\label{eqA}.
\end{eqnarray}
The functions above are
\begin{eqnarray}
& & f_1(\Sigma)= -\frac{\sqrt{2}}{3}\Sigma^2, \;\;\;\qquad  6 f_2(\Sigma)= \frac{4 h}{\Sigma} - l\sqrt{2}\Sigma^2,\;\;\;\qquad f_3(\Sigma)=- l\sqrt{2}\Sigma -\frac{2 h}{\Sigma^2},\nonumber\\
& & f_4(\Sigma)=\frac{\sqrt{2}}{3} \Sigma,\;\;\;\;\;\qquad 6 f_5(\Sigma)= l\sqrt{2}\Sigma +\frac{2h}{\Sigma^2}.\label{fs}
\end{eqnarray}
Let us start from eq.(\ref{eqS}). Solving for $\varphi$, we get
\begin{equation}
e^{2\varphi}=\frac{\Sigma' - f_1}{f_2}\;\;\;\to \;\;2e^{2\varphi} \varphi'= \frac{d}{dr}\left(\frac{\Sigma'- f_1}{f_2}   \right)\ .\label{neweqS}
\end{equation}
Using this in eq.(\ref{eqvarphi}), we obtain
\begin{equation}
 \frac{d}{dr}\left(\frac{\Sigma'- f_1}{f_2}   \right)= f_3 \left(\frac{\Sigma'- f_1}{f_2}   \right)^2.
\end{equation}
We operate with the equation above to obtain
\begin{eqnarray}
\Sigma'' + G_1 \Sigma'(r)^2 + G_2 \Sigma'(r) + G_3=0.\label{eq2nd}
\end{eqnarray}
We have defined
\begin{eqnarray}
 G_1= -\frac{(f_3 +\partial_\Sigma f_2)}{f_2},\;\;\; G_2=\frac{- f_2\partial_\Sigma f_1 + 2 f_1 f_3 + f_1 \partial_\Sigma f_2}{f_2},\;\;\; G_3=- \frac{f_1^2 f_3}{f_2}. \label{Gs}
\end{eqnarray}
Now, we set $\frac{1}{y(\Sigma)}=\Sigma'(r)$, which implies that $\Sigma''(r)= -\frac{1}{y(\Sigma)^3}\partial_\Sigma y(\Sigma)$. This replaced in eq.(\ref{eq2nd}) gives
\begin{equation}
\partial_\Sigma y=  G_1 y + G_2 y^2 + G_3 y^3.\label{eqfinal}
\end{equation}
This is a (first kind) Abel equation in the variable $\Sigma$ for the function $y(\Sigma)$. The functions $G_i$ are given in eqs.(\ref{Gs}) together with (\ref{fs}).
If we solve for $y(\Sigma)$, we can then write
\begin{eqnarray}
& & \dd s^2= e^{2A(\Sigma)}\dd x_{1,3}^2 + y^2(\Sigma){\dd\Sigma^2},\label{esta2}\\
& & e^{2\varphi(\Sigma)}= \frac{1- y(\Sigma) f_1(\Sigma)}{y(\Sigma) f_2(\Sigma)}, \;\;\;\qquad A(\Sigma)= \int \left(f_4(\Sigma) + e^{2\varphi} f_5(\Sigma)\right) y(\Sigma) \dd\Sigma.\nonumber
\end{eqnarray}


\section{The $SO(2)$ invariant truncation of $5D$ $\mathcal{N}=4$ \\$SO(2)\times SE(3)$ gauged supergravity}


This subtruncation retains the modes invariant under the $SO(2)\subset SO(2)\times SE(3)$. The seven remaining scalars are
\begin{equation}
\Sigma,\qquad \mathcal{V}=
\begin{pmatrix}
u & 0 & v\\
0 & 1_{2\times 2} & 0\\
 v & 0 & u 
\end{pmatrix}.
\end{equation}
To make contact with \cite{Cheung:2019pge},  we write
 \begin{equation}
 u = e^{-\varphi_3}U V^{-T} U + e^{\varphi_3}U VU \qquad \mathrm{and} \qquad v = -e^{-\varphi_3}U V^{-T}U+e^{\varphi_3}U VU,
 \end{equation}
 where $U$ is the matrix
 \begin{equation}
 U = \frac{1}{\sqrt{2}}
 \begin{pmatrix}
 0 & 0 & 1\\
 0 & 1 & 0\\
 1 & 0 &0
 \end{pmatrix}.
 \end{equation}
 The submatrices $u$ and $v$ are given by
\begin{equation}
u=
\begin{pmatrix}
\cosh(\varphi_2-\varphi_3) & -\frac{1}{2}a_2e^{\varphi_2-\varphi_3} & -\frac{1}{2}a_3e^{\varphi_2-\varphi_3}\\
\frac{1}{2}a_2 e^{-\varphi_1+\varphi_2 +\varphi_3} & \cosh(\varphi_1-\varphi_2-\varphi_3) & -\frac{1}{2}a_1e^{\varphi_1-\varphi_2-\varphi_3}\\
\frac{1}{2}(a_1 a_2 + a_3)e^{\varphi_1+\varphi_3} & \frac{1}{2}a_1e^{\varphi_1+\varphi_3} & \cosh(\varphi_1+\varphi_3)
\end{pmatrix}
\end{equation}
and
\begin{equation}
v=
\begin{pmatrix}
-\sinh(\varphi_2-\varphi_3) & \frac{1}{2}a_2e^{\varphi_2-\varphi_3} & \frac{1}{2}a_3e^{\varphi_2-\varphi_3}\\
\frac{1}{2}a_2 e^{-\varphi_1+\varphi_2 +\varphi_3} & -\sinh(\varphi_1-\varphi_2-\varphi_3) & \frac{1}{2}a_1e^{\varphi_1-\varphi_2-\varphi_3}\\
\frac{1}{2}(a_1 a_2 + a_3)e^{\varphi_1+\varphi_3} & \frac{1}{2}a_1e^{\varphi_1+\varphi_3} & \sinh(\varphi_1+\varphi_3)
\end{pmatrix}.
\end{equation}

We catalogue the non-vanishing components of some antisymmetric tensors:
\begin{align}
\hat{\xi}^{4 5} & = \, -\sqrt{2}
\end{align}
and defining $\tilde{f}_{m n}\equiv \as_{m n p q r}\hat{f}^{p q r}$,
\begin{align}
\tilde{f}_{4 5} & = 3\big(\mathscr{V}^1\,_1\mathscr{V}^2\,_2\mathscr{V}^8\,_8+\mathscr{V}^1\,_1\mathscr{V}^3\,_3\mathscr{V}^7\,_7+\mathscr{V}^2\,_2\mathscr{V}^3\,_3\mathscr{V}^6\,_6+l\,\mathscr{V}^1\,_1\mathscr{V}^2\,_2\mathscr{V}^3\,_3\nonumber \\ &\quad-\mathscr{V}^2\,_1\mathscr{V}^3\,_3\mathscr{V}^6\,_7-\mathscr{V}^2\,_2\mathscr{V}^3\,_1\mathscr{V}^6\,_8-\mathscr{V}^1\,_1\mathscr{V}^3\,_2\mathscr{V}^7\,_8+\mathscr{V}^2\,_1\mathscr{V}^3\,_2\mathscr{V}^6\,_8\big),
\end{align}
where
\begin{equation}
\mathscr{V} = 
\begin{pmatrix}
e^{-\varphi_3} V^{-T} & 0 & 0\\
0 & 1 & 0\\
0 & 0 & e^{\varphi_3} V
\end{pmatrix}.
\end{equation}
This is sufficient to reproduce the fermion shift matrix, and thus the superpotential appearing in the BPS equations. Note that for the $SO(2)\times SO(2)_R$ and $SO(3)$ invariant sub-tuncations, the second line in $\tilde{f}_{mn}$ vanishes.


\section{The lift to Massive IIA}\label{appendixmassiveiia}


The uplift depends crucially on a class of three-dimensional manifolds the authors of \cite{Passias:2015gya} call $M_3$. This manifold is topologically an $S^3$, but locally an $S^2$ fibration over an interval with coordinate $\zeta$. Associated with the $M_3$ are functions $\mathbb{A}(\zeta), \phi(\zeta)$, and $x(\zeta)$ that obey a system of first order ODE's, see \cite{Passias:2015gya} for these details.

 The metric on the $M_3$ takes the form
\begin{equation}
\dd s_{M_3}^2 = \dd \zeta^2 +\frac{1-x^2}{16 w}e^{2A}D\vec{y}\cdot D\vec{y},
\end{equation}
where $Dy^i \equiv \dd y^i +\as^{ijk} y^j A_{(1)}^k$ with $y^i$ constrained coordinates on $\mathbb{R}^3$ such that $\vec{y}\cdot\vec{y} = 1$, and 
\begin{equation}
w = X^5(1-x^2)+x^2.
\end{equation}
The reader can find a possible representation for the $y^i$ in \cite{Passias:2015gya}.
\\
The ten-dimensional metric is given by
\begin{equation}
\ell^{-1}\dd s^2_{10} = \frac{1}{8}X^{-\frac{1}{2}}e^{2\mathbb{A}}\dd s^2_7+X^\frac{5}{2}\dd s^2_{M_3},
\end{equation}
where $\ell$ is a length parameter which, in our conventions, is $\ell = 8\sqrt{2}$. The dilaton $\Phi$ is
\begin{equation}
e^{2\Phi}= \ell \frac{X^\frac{5}{2}}{w}e^{2\phi}
\end{equation}
and the NS-NS two-form is given by
\begin{equation}
\ell^{-1} B = \frac{1}{16}e^{2\mathbb{A}}\frac{x\sqrt{1-x}}{w}\mathrm{vol}_{\tilde{2}}-\frac{1}{2}e^{\mathbb{A}}\dd\zeta\wj \left(a-\frac{1}{2}\vec{y}\cdot\vec{A}_{(1)}\right),
\end{equation}
where
\begin{equation}
\mathrm{vol}_{\tilde{2}} \equiv \frac{1}{2}\as^{ijk}y^i Dy^{jk}\qquad \mathrm{and} \qquad \dd a = -\frac{1}{2}\mathrm{vol}_{S^2}.
\end{equation}

The R-R fluxes are
\begin{align}
F_2 & = -q\left(\mathrm{vol}_{\tilde{2}}+\vec{y}\cdot\vec{F}_{(2)} \right)+\ell\frac{F_0}{w}e^{2\mathbb{A}}x\sqrt{1-x^2}\mathrm{vol}_{\tilde{2}},\\
\ell^{-1} F_4 & = -\frac{q}{16 w}e^{2\mathbb{A}}x\sqrt{1-x^2}\vec{y}\cdot \vec{F}_{(2)}\wj\mathrm{vol}_{\tilde{2}} -\frac{1}{4}qe^{2\mathbb{A}}\dd\zeta\wj\as^{ijk} F^i_{(2)}\wj y^j Dy^k\nonumber\\
&\qquad -\frac{1}{2}q e^{\mathbb{A}}\dd\zeta\wj X^4 \hf_7 F_{(4)}-\ell^{-1}\frac{1}{2}e^{3\mathbb{A}-\phi}x F_{(4)},
\end{align}
with $F_0$ the Roman's mass and $q=\frac{1}{4}e^{\mathbb{A}-\phi}\sqrt{1-x^2}$.

In the main body of the paper we found it more convenient  to use the variables $(z, \alpha(z))$ defined in \cite{Cremonesi:2015bld}. To define those variables and coordinates, it is first necessary to move to the $(y,\beta(y))$-variables, in terms of which,
\begin{eqnarray}
& & e^{2\mathbb{A}}= \frac{4}{9} \left(-\frac{\partial_y\beta}{y} \right)^{1/2},\;\;\;\;\qquad e^\phi=e^{\phi_0} \frac{ \left(-\frac{\partial_y\beta}{y} \right)^{5/4}}{\sqrt{4\beta -y\partial_y\beta}},\nonumber\\
& & \dd\zeta^2=\frac{1}{9\beta} \left(-\frac{\partial_y\beta}{y} \right)^{3/2}\dd y^2,\;\;\;\;\qquad 1-x^2=\frac{4\beta}{4\beta -y\partial_y\beta}.
\end{eqnarray}
After this change, we move to the variables $(z, \alpha(z))$ of \cite{Cremonesi:2015bld}. The change is,
\begin{eqnarray}
& & \beta= \alpha^2,\;\;\;\qquad y=-\frac{1}{18\pi} \dot{\alpha},\;\;\qquad \dd y= -\frac{1}{18\pi} \ddot{\alpha} \dd z,\nonumber\\
& & \partial_y\beta= -36\pi \frac{\alpha\dot{\alpha}}{\ddot{\alpha}}.\nonumber
\end{eqnarray}
We then express the quantities,
\begin{eqnarray}
& & e^{2\mathbb{A}}=8\pi\sqrt{2} \left(-\frac{\alpha}{\ddot{\alpha}}  \right)^{1/2},\;\;\;\;\qquad \dd\zeta^2=4\pi\sqrt{2} \left(-\frac{\ddot{\alpha}}{\alpha} \right)^{1/2} \dd z^2,\nonumber\\
& & e^{4\phi} =e^{4\phi_0} \frac{\left(-\frac{\alpha}{\ddot{\alpha}}  \right)^3  }{(\dot{\alpha}^2-2\alpha\ddot{\alpha})^2},\;\;\;\qquad 1-x^2= 2\frac{(-\alpha\ddot{\alpha})}{(\dot{\alpha}^2-2\alpha\ddot{\alpha})},\nonumber\\
& & e^{4\Phi_{IIA}}= e^{4\phi}\frac{X^5}{w^2}.\nonumber
\end{eqnarray}
Using these expressions one obtains the values quoted in eq.(\ref{lift10xxx}) for the metric and dilaton. Similarly one can translate the expressions for the Ramond and NS two-form.

It is not particularly enlightening to write explicitly the seven-dimensional fields in terms of those of the five-dimensional theory. The exception is the form of ``admissible" singularities in the ten-dimensional picture. This states that as one approaches the IR of the solution, the quantity
\begin{equation}
|g_{(10)}\,_{tt}|= \frac{1}{8}e^{\varphi}\Sigma^{\frac{1}{2}}e^{2\mathbb{A}}|g_{(5)}\,_{tt}|
\end{equation}
should not increase.

\end{appendix}

 \end{document}